\documentclass[aps,prd,amsmath,amssymb,eqsecnum,nofootinbib,notitlepage,superscriptaddress]{revtex4}

\usepackage[retainorgcmds]{IEEEtrantools}
\usepackage{graphicx}
\usepackage{graphics}
\usepackage{color}

\newtheorem{theorem}{Theorem}

\newtheorem{lemma}{Lemma}

\newcommand{\Mtwo}{\mathcal{M}_2}
\newcommand{\TCSt}{\mathcal{M}_2}
\newcommand{\st}{\mathcal{M}}

\newcommand{\GFret}{G_{R}}

\newcommand{\sTCS}{\sigma}

\newcommand{\be}{\begin{equation}}
\newcommand{\ee}{\end{equation}}

\newcommand{\rs}{r_*}
\newcommand{\lam}{L}
\newcommand{\rsln}{\bar r_*}

\newcommand{\pdd}[2]{\frac{\partial^2{#1}}{\partial{#2}^2}}
\newcommand{\pd}[2]{\frac{\partial{#1}}{\partial{#2}}}
\newcommand{\cgl}{{\cal{G}}_\ell}

\newcommand{\eonetwo}{E_{_{1,2}}}

\newcommand{\ra}{r_1}
\newcommand{\rb}{r_+}
\newcommand{\rc}{r_2}
\newcommand{\ta}{t_1}

\newcommand{\tc}{t_2}
\newcommand{\sqx}{\sqrt{\frac{x+6}{x-2}}}

\begin{document}
\title{
Geometric properties of a 2-$D$ space-time arising in 4-$D$ black hole physics
}

\author{Marc Casals}
\email{mcasals@cbpf.br,marc.casals@ucd.ie}
\affiliation{Centro Brasileiro de Pesquisas F\'isicas (CBPF), Rio de Janeiro, 
Brazil.}
\affiliation{School of Mathematical Sciences and Complex \& Adaptive Systems
Laboratory, University College Dublin, Dublin 4, Ireland.}

\author{Brien C. Nolan}
\email{brien.nolan@dcu.ie}
\affiliation{School of Mathematical Sciences, Dublin City
University, Glasnevin, Dublin 9, Ireland.}

\begin{abstract}
The Schwarzschild exterior space-time is conformally related to a direct product space-time, $\Mtwo \times \mathbb{S}_2$, where $\Mtwo$ is a two-dimensional space-time. This direct product structure arises naturally when considering the wave equation on the Schwarzschild background. Motivated by this, we establish some geometrical results relating to $\Mtwo$ that are useful for black hole physics. We prove that $\Mtwo$ has the rare property of being a \textit{causal domain}. Consequently, Synge's world function and the Hadamard form for the Green function on this space-time are well-defined \textit{globally}. We calculate the world function and the van Vleck determinant on  $\Mtwo$ numerically and point out how these results will be used to establish global properties of Green functions on the Schwarzschild black hole space-time. 
\end{abstract}

\maketitle


\section{Introduction}


In considering the propagation of scalar waves on the Schwarzschild exterior space-time, two observations have significant consequences for any analysis of the problem. 

First is the seemingly mundane observation that one should divide by $r$. That is, in the usual Schwarzschild coordinates, where $r$ is the area radius, technical advantages accrue by considering not the scalar field $\Phi$ itself,  but the rescaled quantity $\Psi$ given by $\Phi=\Psi/r$. (We note that $\Phi$ satisfies the wave equation $\Box_g \Phi = 0$, where $\Box_g$ is the d'Alembertian operator associted with the metric tensor $g_{\alpha\beta}$ of Schwarzschild.) Following a decomposition into spherical harmonic $\ell$-modes, the equation for each mode $\Psi_\ell$ of $\Psi$ then takes the convenient form of a one-dimensional wave equation with an effective potential $V_\ell$ (and a new radial coordinate $\rs$):
\begin{eqnarray}
-\pdd{\Psi_\ell}{t}+\pdd{\Psi_\ell}{\rs}-V_\ell(r)\Psi_\ell = 0.\label{eq:wave-modes}
\end{eqnarray}
Much progress can be made on this basis. See for example Chapter 4 of \cite{Frolov:Novikov}. 

The second observation is that the exterior Schwarzschild space-time has a non-trivial global causal structure. In particular, the exterior Schwarzschild space-time contains circular and other closed null geodesics. This means that pairs of spacetime points may be connected by more than one causal geodesic (e.g.\ a freely falling observer can receive a signal emitted in his or her own past). A technical consequence of this is that a large body of the mathematical machinery developed for the analysis of waves in curved space-time can be applied only to local (in time) results \cite{Friedlander}. For example, neither Synge's world function  (a half of the  geodesic distance squared)  nor the analytic Hadamard form~\cite{Hadamard,DeWitt:1960,Friedlander,Poisson:2011nh} for the Green function associated with the wave equation are globally defined. Similarly, the existence of the photon sphere and trapped null geodesics causes difficulties in the analysis of wave propagation via vector-field and energy methods (see Section 4 of \cite{dafermos2008lectures}). 

The main point of the present paper is that considering a geometric interpretation of the first observation (``divide by $r$") mitigates the difficulties associated in the second observation. The complex causal structure of the Schwarzschild exterior is tightly controlled, and we can make significant progress in identifying globally valid versions of the world function and the Hadamard form for the Green function. The relevant geometric interpretation is the following: the rescaled field $\Psi$ is most naturally understood as a solution of a wave equation of the form $\Box_{\hat{g}}\Psi+u(x^\alpha)\Psi=0$ on the spacetime with metric $\hat{g}$ conformally related to the metric of Schwarzschild via
\be \hat{g}_{\alpha\beta} = r^{-2}g_{\alpha\beta}, \label{eq:conf}\ee
and $u$ is a function on the conformal space-time.

Thus in working with $\Psi$, one is naturally working on the space-time $({\cal{M}},\hat{g}_{\alpha\beta})$, where 
\be {\cal{M}} = \Mtwo \times \mathbb{S}_2 \label{eq:direct-product-mfld}\ee
with (in Schwarzschild coordinates)
\be \Mtwo = \{(t,r):t\in\mathbb{R}, r>2M\}, \label{eq:m2-def}\ee
$M$ is the mass parameter of the Schwarzschild space-time, $\mathbb{S}_2$ is the unit 2-sphere and the line element associated with $\hat{g}_{\alpha\beta}$ is
\be d\hat{s}^2 \equiv r^{-2}ds^2 = -\frac{1}{r^2}(fdt^2-f^{-1}dr^2)+d\Omega_2^2.\label{con-sch-lel}\ee 
Here, $f=1-2M/r$ and $d\Omega_2^2$ is the standard line element of the unit $2$-sphere. We note that this space-time has a direct product structure: the metric tensor takes the form 
\be \hat{g}_{\alpha\beta}(x^\gamma) = \left( \begin{array}{cc} \hat{g}_{AB}(x^C) & 0 \\ 0 & \gamma_{ab}(x^c) \end{array}\right), \label{eq:direct-metric}\ee
where $x^A$ (capital Roman indices) and $x^a$ (lower case Roman) are local coordinates on $\Mtwo$ and $\mathbb{S}_2$ respectively. Space-time coordinates are obtained by concatenation: $x^\alpha=(x^A,x^a)$. Likewise, $\hat{g}_{AB}$ and $\gamma_{ab}$ are metric tensors on $\Mtwo$ and $\mathbb{S}_2$ respectively.
 
We are thus led to the 2-dimensional space-time that is the focus of this paper. We will refer to this as the \textit{2-dimensional conformal Schwarzschild space-time}. This is the space-time $\Mtwo$ with line element 
\be \hat{g}_{AB}dx^Adx^B = ds_2^2 \equiv -\frac{1}{r^2}(fdt^2-f^{-1}dr^2).\label{2d-lel}\ee

Our main result is this:

\begin{theorem}\label{Theorem 1} The spacetime $(\Mtwo,\hat{g}_{AB})$ is a causal domain. 
\end{theorem}

A \textit{causal domain} is a \textit{geodesically convex domain} that is subject to a certain causality condition (see below). A geodesically convex domain is a (region of a) space-time where all pairs of points are joined by a unique geodesic. Being a causal domain is a technically-advantageous property. For example, Synge's two-point world function $\sigma(x,x')$ is defined to be one half of the square of the geodesic distance from $x$ to $x'$. This is well defined only in a geodesically convex domain. Similarly, fundamental solutions of wave equations - such as the analytic Hadamard form~\cite{Hadamard,DeWitt:1960,Friedlander,Poisson:2011nh} - are guaranteed to exist in causal domains (i.e., this is a sufficient, but not necessary, condition for the existence of a fundamental solution). Every point $x$ in a space-time has a neighbourhood which is a geodesically convex domain (\textit{normal neighbourhood}), and so $\sigma$ is always defined locally. For renormalization within quantum field theory on curved space-time, coincidence limits of two-point functions play a central role: thus local existence is sufficient~\cite{Birrell:Davies}.  However, in the study of wave propagation, global existence of $\sigma$ and consequently of the Hadamard form is extremely desirable. Such existence holds in a causal domain. 

The property of a space-time of being a causal domain seems to be an uncommon one. In particular, Schwarzschild space-time is not a causal domain, and so the world function and the Hadamard form are only well-defined in normal neighbourhoods. As a consequence, one must resort to other techniques in order to calculate the Green function outside a normal neighbourhood in Schwarzschild (e.g.,~\cite{CDOW13,Zenginoglu:2012xe,PhysRevD.89.084021,Casals:2012px}).


The main result of this paper is the proof of Theorem 1. In addition, we study the geodesics of $\Mtwo$ and numerically calculate the world-function and the van Vleck determinant, which
are well-defined globally on $\Mtwo$. 
The van Vleck determinant is an important bitensor appearing in the Hadamard form for the Green function which gives a measure of the focusing of the spray of null geodesics emitted from a point~\cite{Visser:1993}. For the Green function presentation and van Vleck determinant results, we focus on the case of a massless scalar field. In a separate paper we use these results in order to obtain a globally-valid expression for the Green function in Schwarzschild space-time and derive some of its properties \cite{casals-nolan-global-hadamard}.

In Sec.\ref{sec:GF} we give details of how the space-time $\Mtwo$ arises naturally when we consider the wave equation of the exterior Schwarzschild space-time, and we describe associated Green functions. 
In Sec.\ref{sec:M2} we recall the definition of a causal domain, we determine properties of the geodesics on $\Mtwo$ and we use these to give a proof of Theorem 1.
In Sec.\ref{sec:sigma_2} we provide the numerical calculation of $\sigma$ and the van Vleck determinant in $\Mtwo$. 
We then make some concluding comments.

Throughout this paper we choose geometric units $c=G=1$ and metric signature $(- + + +)$.


\section{From the wave equation on Schwarzschild to $\Mtwo$}\label{sec:GF}

Our main aim in this section is to show how the space-time $\Mtwo$ arises naturally in the study of the scalar wave equation on the exterior Schwarzschild space-time. The line-element for this space-time can be written as
\be ds^2 = -f(dt^2-d\rs^2) + r^2d\Omega_2^2,\label{sch-lel}\ee
where, as above, $M$ is the mass of the black hole, $f=1-2M/r$ and $d\Omega_2^2$ is the standard line element of the unit $2$-sphere. The tortoise coordinate $\rs$ is related to the area radius $r$ by 
\be \frac{dr}{d\rs}= f.\label{tortoise}\ee 

A massless scalar field $\Phi$ propagating on Schwarzschild space-time satisfies the Klein-Gordon equation
\be \Box_g \Phi =g^{\alpha\beta}\nabla_\alpha\nabla_\beta\Phi = 0.\label{eq:wave-eq}\ee Our principal interest is the study of Green functions for this equation, continuing a line of research that approaches the self-force problem~\cite{Poisson:2011nh} through the study of Green functions on black hole space-times (e.g.,~\cite{CDOW13,PhysRevD.89.084021,Casals:2012px,CDOWa})
(although the global behavior of  Green functions also finds applications in other areas, such as in the calculation of the response of a quantum ``particle
detector"~\cite{Birrell:Davies}). For this reason, we will immediately focus on Green functions rather than on the scalar field itself.

The retarded Green function  $\GFret(x,x')$ is a solution of the inhomogeneous version of (\ref{eq:wave-eq}) with a Dirac-delta distribution as a source, satisfying certain causal boundary conditions: $\GFret(x,x')$ vanishes ouside the past light cone of $x$. In the coordinates above, the wave equation for $\GFret$ reads 
\be -f^{-1}\pdd{\GFret}{t}+f^{-1}\pdd{\GFret}{\rs}+\frac{2}{r}\pd{\GFret}{\rs}+\frac{1}{r^2}\nabla^2\GFret=-\frac{4\pi}{r^2f}\delta_2(x^A-x^{A'})\delta_{\mathbb{S}^2}(x^a,x^{a'}),\label{gfret-eq1}
\ee
where $\nabla^2$ is the Laplacian operator on the unit 2-sphere, $x^A=(t,\rs)$ are coordinates on the Lorentzian 2-space (i.e.\ the 2-space that arises by factoring the 4-dimensional space-time by the action of the $SO(3)$ that generates the spherical symmetry) and $x^a=(\theta,\phi)$ are coordinates on the unit 2-sphere. 

It is standard at this point to rescale the scalar field $\Phi$ by a factor $r$: this removes the first order derivative from the wave equation (\ref{gfret-eq1}) and simplifies the use of a variety of methods from the theory of Partial Differential Equations (e.g.\ WKB approximation methods, frequency domain analysis)\footnote{Note that we could rescale $\Phi(x)$ by any constant multiple of $r$. The corresponding rescaling of $\GFret(x,x')$ likewise involves the factor $r$, multiplied by any term that is constant with respect to the point $x$. For reasons of symmetry, we choose this `constant' multiple to be $r'$: see (\ref{con-gf}).}. It is worth considering the geometrical significance of this step. The appropriate rescaling amounts to making a conformal transformation of the metric and considering $\Psi$ to be the solution of a corresponding equation on the conformal space-time. That is, 
\be \Box_g\Phi = 0 \quad\Leftrightarrow\quad \Box_{\hat{g}}\Psi - \frac{\Box_{\hat{g}}r}{r}\Psi =0,\label{eq:wave-eq-trans} \ee
where $\hat{g}_{\alpha\beta}=r^{-2}g_{\alpha\beta}$. Thus we may consider that the evolution takes place on the spacetime with line element (\ref{con-sch-lel}).

By general properties of Green functions in conformally related space-times, we can write~\cite{Friedlander,Birrell:Davies}
\be \GFret = \frac{1}{r\cdot r'}\hat{G}_R(x,x'),\label{con-gf}\ee
where $\hat{G}_R(x,x')$ is the retarded Green function for the conformally invariant wave equation on the space-time with line element (\ref{con-sch-lel}). (Note that since the Ricci scalar $R$ vanishes in Schwarzschild space-time, the ordinary wave operator $\Box_g$ and the conformally invariant wave operator $\Box_g-R/6$ coincide. Moving to the conformal space-time naturally gives rise to the conformally invariant wave equation on that space-time - which is the second equation of (\ref{eq:wave-eq-trans}).) We will refer to this as the \textit{conformal Schwarzschild space-time}. Using this rescaling, we find 
\be -\pdd{\hat{G}_R}{t}+\pdd{\hat{G}_R}{\rs}+\frac{f}{r^2}(\nabla^2-\frac{2M}{r})\hat{G}_R=-4\pi\delta_2(x^A-x^{A'})\delta_{\mathbb{S}^2}(x^a,x^{a'}).\label{con-gfret-eq1}
\ee

In addition to rendering the wave equation more tractable, the conformal rescaling introduces a very useful simplification of Synge's world function. The world function between two space-time points is half the square of the geodesic distance along a specific geodesic joining the two points~\cite{Hadamard,DeWitt:1960,Friedlander,Poisson:2011nh}. As pointed out earlier, this is well-defined only on regions of space-time in which there exists a unique geodesic between each pair of points. The direct product structure of the metric induced by (\ref{con-sch-lel}) yields 
\be \hat{\sigma}_4 = \sigma(x^A,x^{A'}) + \frac12\gamma^2,\label{world-con}\ee
where $\hat{\sigma}_4$ is the world function of the conformal Schwarzschild space-time and $\sigma(x^A,x^{A'})$ is the world-function of the 2-dimensional Lorentzian space-time with line element (\ref{2d-lel}).
This is the 2-dimensional conformal Schwarzschild spacetime, denoted by $\Mtwo$. We note also that in (\ref{world-con}), $\gamma$ is the geodesic distance on the unit 2-sphere. Then, by conformal invariance of null geodesics, a null geodesic connects $x,x'$ in Schwarzschild space-time if and only if a null geodesic connects the corresponding points of the conformal Schwarzschild space-time.

Next, we examine the space-time geometry of $\Mtwo$. We derive the geodesic equations and some of their properties and use these to prove that  $\Mtwo$ is a causal domain.


\section{Geodesics on $\Mtwo$ and the proof of Theorem 1} \label{sec:M2}

This section gives the main results of the paper. We begin by recalling the definition of a causal domain and showing that $\Mtwo$ satisfies the first of the two relevant criteria. We then write down the geodesic equations of $\Mtwo$ and describe briefly, in terms of these equations, what it is we need to prove in order to establish that $\Mtwo$ is indeed a causal domain. We then give the relevant proof, establishing the required properties of spacelike, null and timelike geodesics in four separate subsections. We note that the argument in the case of timelike geodesics is considerably longer than that for spacelike and null geodesics: issues of uniqueness and existence are dealt with separately. 

\subsection{Causal domains}\label{subsec:cds}

As noted in the Sec.\ref{sec:GF}, the world function of the conformal Schwarzschild space-time decomposes according to (\ref{world-con}). This form is of course only locally valid in the $4$-d space-time, and involves the world function $\sigma(x^A,x^{A'})$ of the 2-dimensional conformal space $\Mtwo$. We will argue here that $\Mtwo$ is a \textit{causal domain} \cite{Friedlander}, so that $\sigma$ is defined \textit{globally}. This yields the major advantage that the Hadamard form for Green functions on $\Mtwo$ (see Eq.(\ref{eq:Had form 2d}) below) is globally valid: this in turn will yield a globally valid expression for the retarded Green function on Schwarzschild space-time~\cite{casals-nolan-global-hadamard}. 

A spacetime $(\st,g_{\alpha\beta})$ is a \textit{causal domain} if 
\begin{itemize}
\item[(CD-i)] any two points $p,q$ of $\st$ are joined by a unique geodesic and 
\item[(CD-ii)] for all pairs of points $p,q$ in $\st$, the set $J^+(q)\cap J^-(p)$ is a compact subset of $\st$, or is empty.
\end{itemize}

The set $J^+(q)$ is the \textit{future emission} of $q\in \st$, defined to be the closure of the chronological future $D^+(q)$ of $q$: 
\be J^+(q)=\overline{D^+(q)},\quad D^+(q)=\{p\in \st:\hbox{ there exists a future-directed time-like geodesic from } q \hbox{ to } p\}.\label{jplus-def}\ee  
The past emission of $q\in \st$, $J^-(q)$, is defined analogously to $J^+(q)$, but with `future-directed' in (\ref{jplus-def}) replaced by `past-directed'.

The space-time of interest is $\Mtwo=(\mathbb{R}^2,g_{AB})$ where in coordinates $x^A=(t,\rs)$, 
\be g_{AB}=\left(\begin{array}{cc} -\frac{f}{r^2} & 0 \\ 0 & \frac{f}{r^2}\end{array} \right). \label{m2-metric}\ee
Then $\Mtwo$ is globally conformally flat: $g_{AB}=r^{-2}f\eta_{AB}$ for all space-time points, where $\eta_{AB}$ is the metric of 2-dimensional Minkowski space-time. It is then straightforward to show that 
\be J^+(q) = \{(t,\rs)\in\mathbb{R}^2: |\rs-\rs(q)|\leq t-t(q), t\geq t(q)\},\label{eq:J+}\ee
and consequently, for any pair of points $p,q \in \Mtwo$, $J^+(q)\cap J^-(p)$ is either a closed diamond, a single point, or is empty. Thus this set is always either compact or empty, and so the condition (CD-ii) for $\Mtwo$ to be a causal domain is satisfied. 

It is less straightforward to show that any two points of $\Mtwo$ are connected by a unique geodesic, but we will prove that this is the case.


\subsection{The geodesic equations}\label{subsec:geos}

\begin{figure}[h!]
\begin{center}
\includegraphics[width=8cm]{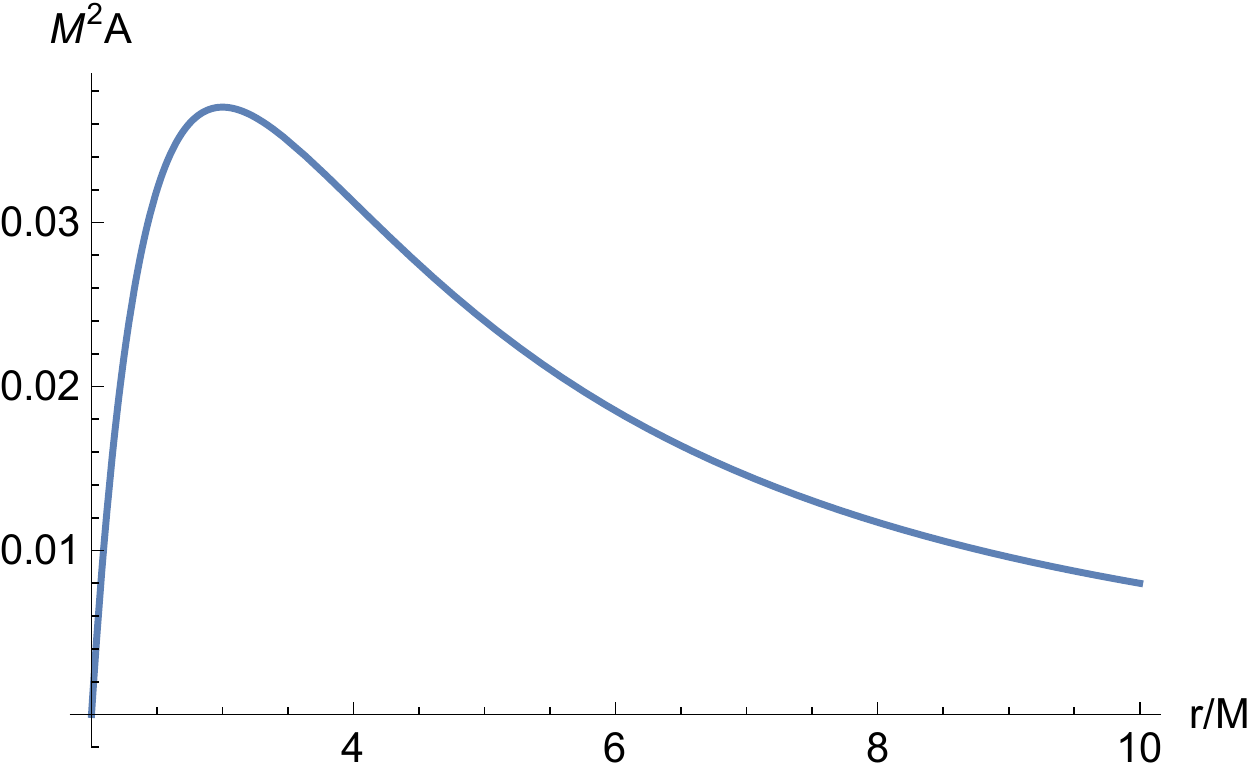}
 \end{center}
\caption{Radial potential $A$ in Eq.(\ref{eq:r dot}) as a function of $r/M$. Its maximum is at $r=3M$.}
\label{fig:Potential}
\end{figure}

In order to study the geodesics on $\Mtwo$, we rewrite the line element as 
\be ds_2^2 = - A(\rs)(dt^2-d\rs^2),\quad A(\rs) = \frac{f}{r^2}.\label{2d-lel-aform}\ee
In (\ref{tortoise}), we choose a constant of integration so that the tortoise coordinate satisfies $\rs=0$ at $r=3M$. Thus 
\be \rs = r-3M +2M\ln\left(\frac{r-2M}{M}\right), \label{rho-def}\ee
and we note that $\rs\to -\infty$ as $r\to 2M^+$, $\rs\to+\infty$ as $r\to+\infty$. Then $A>0$ for all $\rs\in \mathbb{R}$, is increasing on $(-\infty,0)$, is decreasing on $(0,+\infty)$ and has a global maximum of $1/27M^2$ at $\rs=0$ ($r=3M$) -- see the plot in Fig.\ref{fig:Potential}. 
The radii $r=2M$ and $3M$, of course, respectively correspond to the radii of the event horizon and the unstable photon orbit in   Schwarzschild space-time.

The geodesic equations are (with an overdot for derivative with respect to an affine parameter $s$ along the geodesic, and a prime for derivative with respect to argument)
\begin{eqnarray}
\ddot{t}+\frac{A'}{A}\dot{\rs}\dot{t}&=&0,\label{t-geo}\\
\ddot{\rs}+\frac{A'}{A}\dot{\rs}^2-\frac{A'}{2A^2}\epsilon &=&0,\label{ro-geo}
\end{eqnarray}
with the first integral 
\be -A(\dot{t}^2-\dot{\rs}^2) = \epsilon,\label{geo-int}\ee
and where $\epsilon=0,-1,+1$ for null, timelike and spacelike geodesics respectively.

Our aim is to prove that given any pair of points $(t_1,{\rs}_1), (t_2,{\rs}_2) \in \mathbb{R}^2$, there exists a unique value of $\epsilon$ and a unique solution of the equations (\ref{t-geo}), (\ref{ro-geo}) that connects these two points. Without loss of generality, we take $t_1\leq t_2$. 

The equation (\ref{t-geo}) may be integrated to yield
\be A\dot{t} = E,\label{t-con}\ee
where $E$ is constant along the corresponding geodesic. (This conserved quantity corresponds to the existence of the Killing vector $\pd{}{t}$ in $\Mtwo$.) 
Since $A>0$, this means that $t$ is monotone along any geodesic.
Without loss of generality, we choose $E\ge 0$, so
 that $\dot{t}\geq0$ for all geodesics. This amounts to choosing $s$ and $t$ to be co-synchronous along geodesics with $E\neq0$. Furthermore, since $E$ is constant, either $\dot{t}=0$ everywhere along the geodesic, or $\dot{t}>0$ everywhere.

In the case that
$\dot{t}>0$ everywhere along the geodesics in question (we consider the special case $\dot{t}=0$ in the next subsection), we can reduce the problem to one dimension, as we can take $t$ to be the parameter along the geodesic and reduce the geodesic equations to the single equation that arises from (\ref{geo-int}):
\be \left(\frac{d\rs}{dt}\right)^2 = 1 + \epsilon\frac{A}{E^2}.\label{geo-1dim}\ee
The remaining geodesic equation may be taken to be $\dot t=E/A$: given a solution of (\ref{geo-1dim}), this simply serves to determine the parametrisation $t=t(s)$.

In the general case (i.e.\ $\dot{t}\geq 0$), two points $q=(t_1,{\rs}_1)$ and $p=(t_2,{\rs}_2)$ in $\Mtwo$ are \textit{null separated} if $|t_2-t_1|=|{\rs}_2-{\rs}_1|$, \textit{timelike separated} if $|t_2-t_1|>|{\rs}_2-{\rs}_1|$ and \textit{spacelike separated} if $|t_2-t_1|<|{\rs}_2-{\rs}_1|$: notice that any given pair $q,p$ has a unique separation character\footnote{In a general space-time,  a pair of points $p,q$ are null (timelike, spacelike) separated if there is an everywhere null (timelike, spacelike) curve that connects the points. Pairs of points do not necessarily have a unique separation character: consider e.g.\ points $(t,r)=(t_1,3M)$ and $(t,r)=(t_2,3M)$ on the photon sphere $r=3M$ in Schwarzschild space-time, with $t_1,t_2$ chosen so that a circular null geodesic at $r=3M$ passes through both points. These are also connected by a time like curve with $r=3M$ and $\theta,\phi$ constant. However, since $A(\rs)>0$ for all $\rs\in\mathbb{R}$, it is straightforward to show that the separation character is unique in $\Mtwo$ and that it may be described as stated in the text.}. For geodesics with $\dot{t}>0$, it is clear from (\ref{geo-1dim}) that 
\be \left|\frac{d\rs}{dt}\right| \left\{ \begin{array}{c} =1, \\ <1, \\ >1, \end{array}\right. \Leftrightarrow
\epsilon \left\{ \begin{array}{c} = 0, \\ <0, \\ > 0. \end{array}\right.
\label{eq:null/tlike/slike}
\ee
It is also clear from Eq.(\ref{geo-int}) that the case $\dot t=0$ and $\dot \rs\neq 0$ corresponds to $\epsilon>0$, i.e., a spacelike geodesic.

It follows that null (respectively timelike; spacelike) separated points can be connected only by null (respectively timelike; spacelike) geodesics. This elementary observation allow us to decompose the proof that (CD-i) holds in a useful way, by considering the pair of points $p_1=(t_1,{\rs}_1), p_2=(t_2,{\rs}_2)$ from the point of view of their separation character. That is we prove that if $|t_2-t_1|=|{\rs}_2-{\rs}_1|$, then there is a unique null geodesic from $p_1$ to $p_2$; if $|t_2-t_1|<|{\rs}_2-{\rs}_1|$, then there is a unique spacelike geodesic from $p_1$ to $p_2$ and finally if $|t_2-t_1|>|{\rs}_2-{\rs}_1|$, then there is a unique timelike geodesic from $p_1$ to $p_2$. This exhausts all possibilities and so establishes (CD-i). 

Before proceeding, we note that we can relate the radial coordinate $r$ and the affine parameter $s$ by combining Eqs.(\ref{t-con}) and (\ref{geo-int}) to obtain
\begin{align}
\frac{\dot{r}^2}{r^4}=E^2+ \epsilon A,
\label{eq:r dot}
\end{align}
with $A=A(\rs)$ acting as a radial potential.
We plot this potential in Fig.\ref{fig:Potential}.

\subsection{Null separations - existence and uniqueness}\label{subsec:null-sep}
It is trivial to show that null separated pairs of points are connected by a unique null geodesic. The required geodesic is 
\be \rs(t) = {\rs}_1 + \frac{{\rs}_2-{\rs}_1}{t_2-t_1}(t-t_1).\ee

\subsection{Spacelike separations  - existence and uniqueness}\label{subsec:space-sep}

We have two cases to deal with here: $t_2=t_1$ and $t_2>t_1$. In the former case, we must have $\dot{t}=0$ everywhere along the geodesic (since $\dot{t}\geq0$). Then $\epsilon =+1$ in (\ref{geo-int}) and we can integrate to obtain the general form 
\be r(s) = M(1+\cosh (s-s_0)), \ee
where $s$, the affine parameter along the geodesic, represents (non-dimensionalized) proper distance. Initial and terminal values of $s$ then yield the unique geodesic connecting $(t_1,r_1)$ and $(t_1,r_2)$.

For the case $\dot{t}>0$, we may take ${\rs}_2>{\rs}_1$ without loss of generality. Then we seek a unique value of $E$ and a unique solution of 
\be \frac{d\rs}{dt} = \sqrt{1+\frac{A}{E^2}}\label{geo-spacelike}\ee
that connects these points. For a given value of $E$ and a given initial point $q=(t_1,{\rs}_1)$, we can formally write down the solution of (\ref{geo-spacelike}):
\be t_2(E)-t_1 = \int_{{\rs}_1}^{{\rs}_2} \frac{E}{\sqrt{E^2+A(\rs)}}d\rs.\label{t2-E}\ee
As emphasised, this equation yields the value of $t_2$ as a function of $E$ with $t_1,{\rs}_1$ and ${\rs}_2$ taken to be fixed. We can then calculate
\be \frac{dt_2}{dE} = \int_{{\rs}_1}^{{\rs}_2} \frac{A(\rs)}{(E^2+A(\rs))^{3/2}}d\rs >0,\ee
so that the function $E\mapsto t_2(E)$ is monotone increasing. We note also that $t_2(0)=t_1$. For any given pair of values ${\rs}_1,{\rs}_2$, the function $A(\rs)$ has strictly positive lower and upper bounds on $[{\rs}_1,{\rs}_2]$. It follows that 
\be \lim_{E\to +\infty}\frac{E}{\sqrt{E^2+A(\rs)}} = 1\ee
uniformly on $[{\rs}_1,{\rs}_2]$, and so for ${\rs}_1,{\rs}_2$ fixed,
\begin{eqnarray} \lim_{E\to +\infty} t_2(E) - t_1 &=&\lim_{E\to+\infty} \int_{{\rs}_1}^{{\rs}_2} \frac{E}{\sqrt{E^2+A(\rs)}}d\rs \nonumber \\ 
&=& \int_{{\rs}_1}^{{\rs}_2} d\rs\nonumber \\
&=& {\rs}_2-{\rs}_1.
\end{eqnarray}
It follows that $E\mapsto t_2(E)-t_1$ is a one-to-one mapping of $(0,\infty)$ onto $(0,{\rs}_2-{\rs}_1)$. Therefore given any pair of points satisfying the spacelike separation condition $t_2-t_1<{\rs}_2-{\rs}_1$, there exists a unique value of $E$ such that (\ref{t2-E}) provides the unique spacelike geodesic connecting these two points.

\subsection{Timelike separations - general issues and uniqueness}\label{subsec:time-sep}

Let us now consider the case of timelike geodesics.
This case is more complex than the others, as the right hand side of (\ref{geo-1dim}) can vanish along the geodesic, depending on the value of $E$. Recalling that $A(\rs)$ has a global maximum of $\frac{1}{27M^2}$ at $\rs=0$, we define the critical value of $E$ to be 
\be E_c = \frac{1}{3\sqrt{3}M}.\label{e-crit}\ee

This gives rise to the following classification of timelike geodesics:
\begin{itemize}
\item[(i)] \textbf{Supercritical timelike geodesics.} These have $E>E_c$, and so - from (\ref{geo-1dim}) - $\left(\frac{d\rs}{dt}\right)^2>0$ for all $t$. Hence $t\mapsto \rs(t)$ is monotone along these geodesics. As the right-hand side of (\ref{geo-1dim}) is uniformly bounded and is bounded away from zero, these geodesics are defined for all $t\in\mathbb{R}$ and satisfy either $\lim_{t\to\pm\infty}\rs(t)=\pm\infty$ or $\lim_{t\to\pm\infty}\rs(t)=\mp\infty$ depending on the sign of $\frac{d\rs}{dt}$. The supercritical geodesics overcome the potential barrier at $\rs=0$.  
\item[(ii)] \textbf{Critical timelike geodesics.} These have $E=E_c$, and (\ref{geo-1dim}) may be written in the form
\be \left(\frac{d\rs}{dt}\right)^2 = 1 -\frac{A(\rs)}{A(0)}.\label{geo-crit}\ee
The unique geodesic for which $\rs(t_0)=0$ at some $t_0\in\mathbb{R}$ is the static geodesic $\rs(t)\equiv 0$. Thus all other critical geodesics are confined to either $\rs<0$ or $\rs>0$. These satisfy either $\lim_{t\to-\infty}\rs(t)=0$ or $\lim_{t\to+\infty}\rs(t)=0$. 
\item[(iii)] \textbf{Subcritical timelike geodesics.} These have $E<E_c$. For a given $E\in(0,E_c)$, there exist unique values ${\rs}_\pm(E)$, with ${\rs}_+(E)>0$ and ${\rs}_-(E)<0$, for which $1-A({\rs}_\pm)/E^2=0$. These values correspond to turning points of the geodesics: differentiating (\ref{geo-1dim}) and noting that $\rs'(t)\not\equiv0$ yields
\be \frac{d^2\rs}{dt^2} = - \frac{A'(\rs)}{2E^2},\label{ropp}\ee
so that subcritical geodesics in $\rs>0$ have a global minimum at ${\rs}_+(E)>0$, and subcritical geodesics in $\rs<0$ have a global maximum at ${\rs}_-(E)<0$. These geodesics reflect off the potential barrier at ${\rs}_\pm$. We note also that for a subcritical geodesic with $\rs>0$ (respectively, $\rs<0$), there is a unique value $t_0$ such that $\rs'(t)<0$ for $t<t_0$, $\rs(t_0)={\rs}_+$ and $\rs'(t)>0$ for $t>t_0$, and $\rs(t)\to+\infty$ as $t\to\pm\infty$ (respectively, $\rs'(t)>0$ for $t<t_0$, $\rs(t_0)={\rs}_-$ and $\rs'(t)<0$ for $t>t_0$, and $\rs(t)\to-\infty$ as $t\to\pm\infty$). 
\end{itemize}

The simple observation that 
\be 0<E_1<E_2 \quad\Rightarrow\quad 1 - \frac{A(\rs)}{E_1^2} < 1 - \frac{A(\rs)}{E_2^2} \label{E-ineq}\ee
allows us to rule out multiple crossings, and thus demonstrate uniqueness, for many cases of pairs of geodesics. For example, for a pair of supercritical geodesics with $E_c<E_1<E_2$, both satisfying $\frac{d\rs}{dt}>0$, this shows that (with obvious notation) $\frac{d{\rs}_1}{dt}<\frac{d{\rs}_2}{dt}$ at a point of intersection. Thus $\rs={\rs}_1(t)$ always crosses $\rs={\rs}_2(t)$ from above, and so the geodesics meet at most once. Variations on this argument immediately rule out multiple intersections except for the case of a pair of subcritical geodesics.

We will not go through all these possible variations, but by way of example, deal with one case, that of a supercritical or critical timelike geodesics meeting a subcritical timelike geodesic twice. For example, consider a critical or supercritical geodesic (so that $E=E_2\geq E_c$) with $\frac{d\rs}{dt}<0$ and a subcritical geodesic confined to $\rs>0$ (and with $E=E_1<E_c$). By (\ref{E-ineq}), the supercritical geodesics  either (a) crosses the subcritical geodesic from above at some $\rs<{\rs}_+$, the minimum of $\rs$ on the subcritical geodesic: a second crossing could only occur on the decreasing branch of the subcritical geodesic, but this is ruled out by comparison of slopes; or (b) first crosses the subcritical geodesic on its increasing branch at some $\rs\geq{\rs}_+$, and no other crossing is possible. 

Thus, to establish uniqueness, it remains to consider the case of two subcritical geodesics intersecting twice. This corresponds to a pair of particles $P_A$ and $P_B$, moving on timelike geodesics $\gamma_A$ and $\gamma_B$ with energies $E_A<E_B<E_c$, which meet at the points $({\rs}_1,t_1)$ and $({\rs}_2,t_2)$ with $t_2>t_1$. We argue as follows that this is not possible. 

The argument is quite straightforward, but has some technicalities (relating to the second and third points listed below) that are detailed below. We consider the case of two subcritical timelike geodesics confined by the potential barrier to $\rs>0$. The case $\rs<0$ is similar. 

The uniqueness argument is based on the following observations:
\begin{itemize}
\item[O-1.] The relation between the energies and slopes of the trajectories at a point of intersection is constrained by (\ref{E-ineq}).
\item[O-2.] The motion of the particle is symmetric about the minimum. That is, $\rs(t_+-t)=\rs(t_++t)$ for all $t\in\mathbb{R}$, where the minimum $\rs={\rs}_+$ occurs at $t=t_+$. We refer to $t_+$ as the \textit{arrival time}.
\item[O-3.] The minimum value of $\rs$ along a subcritical geodesic is a decreasing function of $E$. That is, ${\rs}_+'(E)<0$. See  (\ref{rplus-E-decr}) - recalling that $\frac{d\rs}{dr}>0$.
\item[O-4.] Consider all subcritical geodesics, indexed by $E<E_c$, which decrease through a point $(t_1,{\rs}_1)$. Define $\Delta_1(E) = t_+(E)-t_1$, where $t_+(E)$ is the \textit{arrival time} of geodesic $E$, i.e.\ the time at which geodesic $E$ meets its minimum. Then $\Delta_1'(E)>0$, so that lower energy particles meet their minimum before higher energy particles emitted from the same point at the same time. Likewise, subcritical geodesics which increase through a point $(t_2,{\rs}_2)$ satisfy $\Delta_2'(E)>0$, where $\Delta_2(E)=t_2-t_+(E)$.
\end{itemize} 

\begin{figure}
		{\includegraphics{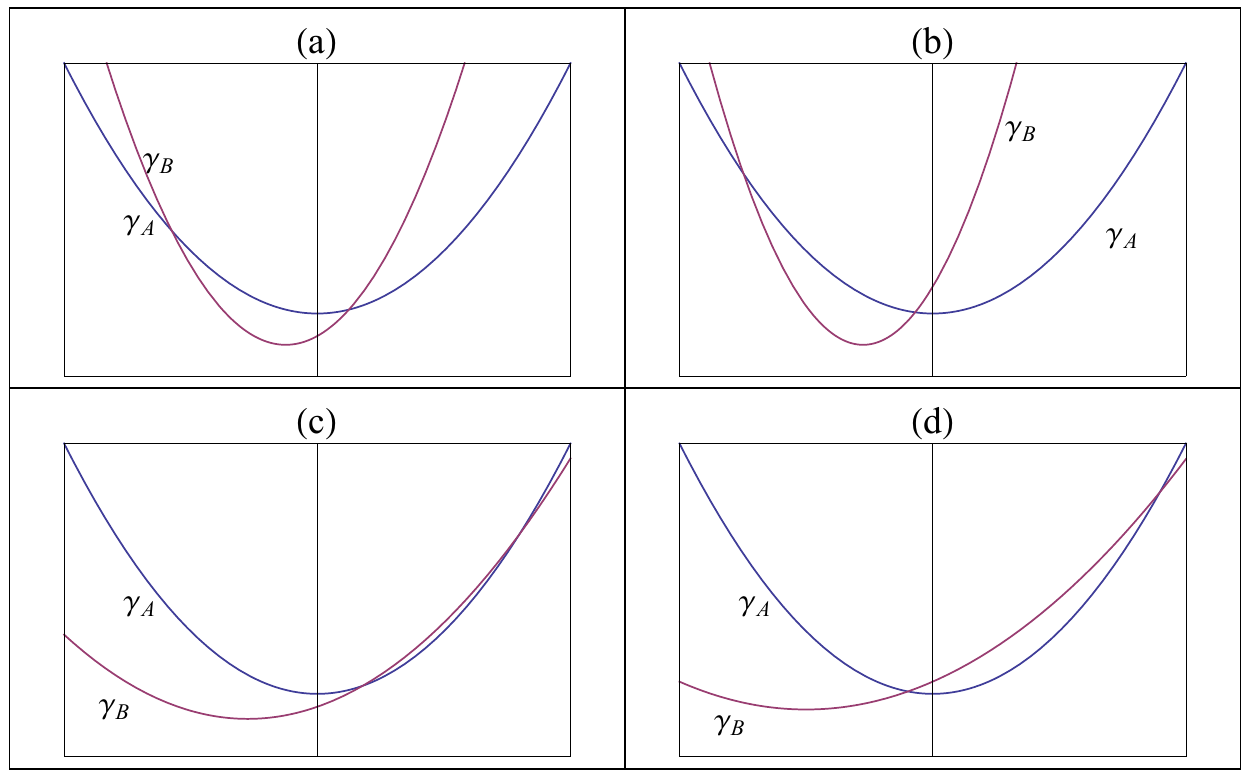}} 
		\caption{The figure shows the four inequivalent ways in which a pair of subcritical timelike geodesics could intersect. The diagrams are representative - our aim is to prove that such intersections cannot occur - but note that O-2 is respected in the diagrams. In each graph, $\rs$ runs along the vertical axis and $t$ on the horizontal, and  $E_A<E_B$ in each case ($E_i$ being the value of $E$ along the geodesic $\gamma_i$. Notice that this gives rise to a lower minimum for $\gamma_B$. By a translation, we fix the minimum of $\gamma_A$ to be at $t=0$.}
\label{fig:four-crossings}
\end{figure}

Subject to these conditions, there are then four possible ways in which the two subcritical geodesics can meet twice, illustrated in Figure~\ref{fig:four-crossings}. 
 
We rule out these four cases as follows. 

\begin{itemize}
\item[(a)] The geodesics intersect at $(t_1,{\rs}_1)$ and $(t_2,{\rs}_2)$ with $t_1<0<t_2$. By O-4 above, 
\be  \Delta_1(E_A) = t_+(E_A)-t_1 < t_+(E_B) - t_1 = \Delta_1(E_B),\ee
since $E_A<E_B$ (this is correctly represented in the diagram).
But also 
\be \Delta_2(E_A) = t_2 - t_+(E_A) < t_2 - t_+(E_B) = \Delta_2(E_B).\ee
Adding the inequalities leads to the contradiction $t_2-t_1<t_2-t_1$, and so case (a) is ruled out. 
\item[(b)] Case (b) also contradicts O-4, as $\gamma_B$ - the higher energy geodesic - reaches its minimum before $\gamma_A$ does, having started from a common point. 
\item[(c)] In case (c), the second intersection point violates the slope inequality of O-1: on an increasing branch, the higher energy geodesic $\gamma_B$ can only meet the lower energy geodesic $\gamma_A$ from below.
\item[(d)] The same argument as case (c) applies to rule out case (d). 
\end{itemize}  

To complete the argument, it remains to establish the claims of O-2, O-3 and O-4. 

The first of these is trivial: each branch of the geodesic reaching/emerging from the minimum is the unique solution of the relevant time-reversed initial value problem. The condition (O-3) follows immediately from (\ref{rplus-E-decr}) below: establishing (O-4) requires more effort. 

So consider a subcritical geodesic with energy $E<E_c$ (recall that $E_c$ is the maximum value of the energy $E$ such that the timelike geodesic bounces off the potential)
 with initial value ${\rs}_1=\rs(t_1)>0$, and which is initially decreasing, so that 
\be \left.\frac{d\rs}{dt}\right|_{t=t_1} = -\sqrt{1-\frac{A({\rs}_1)}{E^2}}.\label{rho-init}\ee
That is, the subcritical geodesic starts off at an initial point $(\ta,\ra)$ and bounces off the potential barrier at the periapsis $\rs ={\rs}_+$. For this case we must have $\sqrt{A({\rs}_1)}<E<E_c$.

The value of $\rs$ along the geodesic decreases monotonically until it reaches its global minimum $\rs={\rs}_+(E)$ at time $t=t_+(E)$. A formal integration yields
\be \Delta_1(E) = t_+(E) - t_1 = \int_{{\rs}_+(E)}^{{\rs}_1} \frac{E d\rs}{\sqrt{E^2-A(\rs)}},\label{t-plus-int1}\ee
or, returning for convenience to the area radius coordinate $r$,
\be \Delta_1(E) = \int_{r_+(E)}^{r_1}
\frac{1}{\sqrt{1-\frac{1}{E^2r^2}\left(1-\frac{2M}{r}\right)}}\frac{dr}{\left(1-\frac{2M}{r}\right)}.
\label{t-plus-int2}\ee
Here,  $r_+(E)$ is the unique root of 
\be E^2r^3-r+2M=0\label{r-plus-eqn}\ee
with $r>3M$. With $E<E_c$, (\ref{r-plus-eqn}) has three real roots; one negative root,  denoted $r_n$, 
and two positive roots: $r_-\in(2M,3M)$ and $r_+\in(3M,+\infty)$.
We note that the area radii $r_{\pm}$ correspond to the tortoise radii ${\rs}_{\pm}$.
 Implicit differentiation of (\ref{r-plus-eqn}) yields
\be \frac{dr_+}{dE}=-Er_+^3\left(1-\frac{3M}{r_+}\right)^{-1}<0.\label{rplus-E-decr}\ee
This inequality has a simple physical interpretation: subcritical particles with higher energy penetrate deeper into the potential barrier. Note that this establishes O-3.

Monotonicity allows us to consider $t_+$ as a function of $r_+$, and substituting for $E^2$ from (\ref{r-plus-eqn}) yields
\be \delta_1(r_+) =   \int_{r_+}^{r_1} \frac{r^{3/2}\left(1-\frac{2M}{r}\right)^{-1}dr}{\left[(r-r_+)(r^2+r_+r-\frac{2M{r_+}^2}{r_+-2M})\right]^{1/2}},\label{t-plus-r-plus}
\ee
where $\delta_1(r_+)=\Delta_1(E)$. As the particle increases away from the minimum, it reaches ${\rs}_2$ at time $t_2>t_+>t_1$ and we can write
\be \Delta_2(E) = t_2 - t_+(E) = \int_{{\rs}_+}^{{\rs}_2}\frac{E d\rs}{\sqrt{E^2-A(\rs)}},\label{t-plus-int3}\ee
or equivalently (with $\delta_2(r_+)=\Delta_2(E)$)
\be \delta_2(r_+) =  \int_{r_+}^{r_2} \frac{r^{3/2}\left(1-\frac{2M}{r}\right)^{-1}dr}{\left[(r-r_+)(r^2+r_+r-\frac{2M{r_+}^2}{r_+-2M})\right]^{1/2}}.\label{t-plus-r-plus2}
\ee

To establish O-4 above, it remains to show that for any fixed $x_0>x>3M$, the function
\be x\mapsto \delta_t(x)= \int_x^{x_0} \frac{r^{3/2}\left(1-\frac{2M}{r}\right)^{-1}dr}{\left[(r-x)(r^2+xr-\frac{2M{x}^2}{x-2M})\right]^{1/2}}\label{delta-x}
\ee
is decreasing. This follows from the fact that (taking $x_0=r_1$ and $x=r_+$)
\be \Delta_1'(E) = \delta_t'(r_+)\frac{d r_+}{dE},\ee
and that $\frac{d r_+}{dE}<0$, as we know from (\ref{rplus-E-decr}). The case for $\Delta_2(E)$ is identical. 

By rescaling, and without loss of generality, we set $M=1$ for the remainder of this subsection.
If we introduce 
\be x = r_+, \label{eq:x-def}\ee
then considering the nature of the roots of the cubic equation (\ref{r-plus-eqn}) allows us to write
\be r_-=b(x),\quad r_n = -x-b(x).\label{eq:roots-xb}\ee
where 
\be b(x) = \frac{x}{2}\left(-1+\sqrt{\frac{x+6}{x-2}}\right).\label{b-def}\ee
We note that 
\be b(x) < x \quad \hbox{ for all } x>3. \label{eq:b-x}\ee
Then we can write
\be \delta_t(x) = \int_x^{x_0} \frac{r^{5/2}dr}{(r-2)\sqrt{r-x}\sqrt{r-b(x)}\sqrt{r+x+b(x)}}.\label{delta-new}\ee

Thus to prove that $\Mtwo$ is a causal domain, we must show that $\delta$ is a decreasing function of $x$ on the interval $(3,x_0)$, for all $x_0>3$. 



Let us recall the meaning of $\delta_t(x)$. For each fixed $x_0$, this measures the elapse of coordinate time $t$ as a particle falls from $r=x_0$ to its minimum $r=x=r_+$ (which by symmetry, is the same as the time to reflect off the potential barrier at $r=x$ and move outwards to $r=x_0$). Instead of dealing with the coordinate time $t$, we shall find it easier to deal with the proper time $\tau$.

It is straightforward to show that the function corresponding to $\delta_t$ is 
\be \delta_\tau(x) = \frac{1}{E(x)}\int_x^{x_0} \frac{dr}{\left[r(r-x)(r-b)(r+x+b)\right]^{1/2}},
\label{eq:del-tau-def}\ee
and the problem now becomes to prove that $x\mapsto\delta_\tau(x)$ is a decreasing function of $x$ on ($3,x_0)$ for all $x_0>3$. Note that we have emphasized that $E$ is a function of $x$. Indeed since $r=x=r_+$ solves (\ref{r-plus-eqn}), we have 
\be E(x) = \left( \frac{x-2M}{x^3}\right)^{1/2}.\label{eq:E-x}\ee
The advantage of moving to the proper time representation is that $\delta_\tau$ may be written in terms of a familiar special function (see Eq. (3.147-8) of \cite{GradRyz}:
\be
\delta_\tau(x) = \alpha F(\psi,q) \label{eq:deltau}\ee
where 
\begin{equation}
F(\psi,q)=\int_0^{\psi}\frac{d\theta}{\sqrt{1-q\sin^2\theta}} \label{eq:elliptic-f}
\end{equation}
is  Legendre's incomplete elliptic integral of the first kind~\cite{Mathematica}
\footnote{Here and below, we follow  {\rm MATHEMATICA}'s~\cite{Mathematica} definition of the elliptic integrals, which differs slightly from that in~\cite{GradRyz}.} and 
\begin{eqnarray}
\alpha & \equiv & \frac{2}{E\sqrt{(r_--r_n)r_+}} = \left(\frac{2b(x+b)}{x+2b}\right)^{1/2},\label{eq:alpha}\\ 
\psi & \equiv&  \sin ^{-1} \bar\varphi>0,\quad \bar\varphi\equiv \sqrt{\varphi},\label{eq:phi-def}\\
\varphi & \equiv &  \frac{(r_--r_n)   (x_0-r_+)}{ (r_+-r_n)(x_0-r_-)} = \frac{(x+2b)(x_0-x)}{(2x+b)(x_0-b)}, \label{eq:varphi-def}\\
q & \equiv & \frac{r_- (r_+-r_n)}{r_+(r_--r_n)} = \frac{b(2x+b)}{x(x+2b)}.\label{eq:q-def}
\end{eqnarray}

It is straightforward to establish that 
\be 0< \varphi < 1,\quad 0 < \psi < \pi/2, \quad 0 < q < 1 \qquad \hbox{ for all } x_0 > x> 3. \label{eq:bounds} \ee

We shall now use the exact result in Eq.(\ref{eq:deltau}) to prove that $\delta_\tau$ is a decreasing function of its argument, and so prove O-4.

We note the partial derivatives
\begin{eqnarray}
\pd{F(\psi,q)}{\psi} &=& \frac{1}{\sqrt{1-q\sin^2\psi}},\label{eq:ellipticf-pd1}\\
\pd{F(\psi,q)}{q} &=& \frac{1}{2q(1-q)}\left(E(\psi,q)-(1-q)F(\psi,q)-q\frac{\cos\psi\sin\psi}{\sqrt{1-q\sin^2\psi}}\right), \label{eq:ellipticf-pd2}
\end{eqnarray}
where 
\be E(\psi,q) = \int_0^\psi \sqrt{1-q\sin^2\theta}d\theta \label{eq:elliptic-e} \ee
is Legendre's incomplete elliptic integral of the first kind. Then we find, using (\ref{eq:phi-def}),
\begin{eqnarray}
\delta_\tau'(x) &=& \frac{\alpha}{2(1-q)\sqrt{1-q\varphi}\sqrt{\varphi}\sqrt{1-\varphi}}\left((1-q)\varphi'-q'\varphi(1-\varphi)\right) \nonumber \\
&& + \left(\alpha'-\frac{\alpha}{2}\frac{q'}{q}\right)F(\psi,q)+\frac{\alpha q'}{2q(1-q)}E(\psi,q). \label{eq:deltau-prime}
\end{eqnarray}

To establish the decreasing nature of $\delta_\tau$, we show that 
\be {\cal{T}}_1 = (1-q)\varphi'-q'\varphi(1-\varphi) \label{eq:ct1} \ee
and
\be {\cal{T}}_2 = \beta F(\psi,q) + \gamma E(\psi,q) \label{eq:ct2} \ee
are both negative for all $x\in(3,x_0)$ and all $x_0>3$. Note that we have introduced
\be \beta = \alpha'-\frac{\alpha}{2}\frac{q'}{q},\qquad \gamma = \frac{\alpha q'}{2q(1-q)}. 
\label{eq:beta-gamma}
\ee

For ease of reading, we separate these results into two lemmas. 

\begin{lemma} ${\cal{T}}_1 <0$ for all $x\in(3,x_0)$ and all $x_0>3$. 
\end{lemma}
\noindent{\textbf{Proof:}} A straightforward calculation shows that 
\be {\cal{T}}_1 = \frac{-16x_0}{(x-2)^3\sqx(3+\sqx)^2(2b-2x_0)^2}\times(u+\sqx v),\ee
where
\begin{eqnarray}
u &=& (x-3)^3+5(x-3)^2+5(x_0-3)(x-3)+9(x_0-3),\\
v &=& (x-2)((x-3)^2-3(x_0-3)).
\end{eqnarray}
It is immediate that $u>0$ for all $x\in(3,x_0)$ and all $x_0>3$. If $(x-3)^2\geq 3(x_0-3)$, then $v$ is non-negative, and ${\cal{T}}_1$ is negative, proving the lemma. If $(x-3)^2<3(x_0-3)$, then ${\cal{T}}_1$ is negative if and only if 
\be u > -\sqx v \quad \Leftrightarrow \quad u^2 > \left(\frac{x+6}{x-2}\right) v^2. \ee Expanding terms shows that this last inequality is equivalent to 
\be (34+16(x_0-3))(x-3)^4 + 128(x_0-3)(x-3)^3+(16(x_0-3)^2+144(x_0-3))(x-3)^2 > 0, \ee
which is clearly true for all $x\in(3,x_0)$ and all $x_0>3$. This completes the proof. \hfill $\blacksquare$

\begin{lemma}  ${\cal{T}}_2 <0$ for all $x\in(3,x_0)$ and all $x_0>3$. 
\end{lemma}
\noindent{\textbf{Proof:}} From the definitions (\ref{eq:elliptic-f}) and (\ref{eq:elliptic-e}), we can write
\be {\cal{T}}_2 = \int_0^\psi \frac{\beta+\gamma-\gamma q \sin^2\theta}{\sqrt{1-q\sin^2\theta}}d\theta. \ee
We find that 
\be \gamma q = \frac{\alpha q'}{2(1-q)} < 0 \quad \hbox{ for all } x>3, \ee
and so 
\be \beta + \gamma - \gamma q \sin^2\theta < \beta + \gamma -\gamma q \sin^2\psi, \quad \theta\in(0,\psi). \ee
Using (\ref{eq:phi-def}), this allows us to write 
\be {\cal{T}}_2 < (\beta + \gamma(1-q\varphi)) F(\psi,q), \ee
and so the lemma is proven if we can establish the inequality
\be \beta + \gamma(1-q\varphi) = \frac{\alpha}{2}\left[ \frac{(\alpha^2)'}{\alpha^2}+\left(\frac{1-\varphi}{1-q}\right)q'\right] < 0 \quad \hbox{ for all } x\in(3,x_0), x_0>3. \ee
We find that
\be \frac{(\alpha^2)'}{\alpha^2} + \left(\frac{1-\varphi}{1-q}\right)q' = 
\frac{-x^2}{(x-2)^3{\sqx}b(x+b)(2x+b)(x_0-b)}\times (y + \sqx z), \ee
where
\begin{eqnarray} 
y &=& (x_0+1)(x-3)^2+5x_0(x-3)+3(4x_0-3),\\
z &=& -((x_0-3)(x-3)-9)(x-2).
\end{eqnarray}
We see immediately that $y>0$ for all $x\in(3,x_0), x_0>3$. Thus ${\cal{T}}_2<0$ if $z\geq 0$. If $z<0$, then
\be y+\sqx z > 0\quad \Leftrightarrow \quad y^2 > \left(\frac{x+6}{x-2}\right) z^2. \ee Expanding terms shows that this last inequality is equivalent to 
\be (x_0-1)(x-3)^4 + (11x_0-18)(x-3)^3 + ((x_0-3)^2+12(x_0-3)+9)(x-3)^2+3(x_0-3)(5x_0+18)(x-3)+9(x_0-3)(2x_0-3) > 0 \ee
which is clearly true for all $x\in(3,x_0)$ and all $x_0>3$. This completes the proof of the lemma. \hfill $\blacksquare$

This completes the proof that timelike separated pairs of points in $\Mtwo$ are connected by at most one timelike geodesic. We complete the proof of Theorem 1 in the following subsection by proving that timelike separated pairs of points are connected by at least one timelike geodesic.

\subsection{Timelike separations - existence}

We consider two points $(t_1,{\rs}_1)$, $(t_2,{\rs}_2)$ which satisfy the time-like separation condition
\be |{\rs}_2-{\rs}_1| < |t_2-t_1| = t_2-t_1,\label{time-sep}\ee
where we impose, without loss of generality, $t_1<t_2$. Our aim is to prove that there exists a value $\eonetwo$ of $E$ and a solution of 
\be \left(\frac{d\rs}{dt}\right)^2 = 1-\frac{A(\rs)}{E^2} \label{t-geo-ex}\ee
which satisfies $\rs(t_1)={\rs}_1$ and $\rs(t_2)={\rs}_2$. In addition, we specify that ${\rs}_1<0$ (corresponding to $r(t_1)<3M$): the arguments carry over, mutatis mutandi, to the case ${\rs}_1\geq 0$.  


We introduce $\rs=\rsln(t)$, the solution of the initial value problem (IVP)
\be \left(\frac{d\rs}{dt}\right)^2 = 1-\frac{A(\rs)}{A({\rs}_1)},\qquad \rs(t_1)={\rs}_1. \label{rho-star}\ee
That is, this geodesic has energy $E_*=\sqrt{A({\rs}_1)}$. This geodesic exists and is uniquely defined on $\mathbb{R}$. It has the distinguishing property that $\rs=\rsln(t)$ has a global maximum of $\rs={\rs}_1$ which is attained at $t=t_1$. Thus this geodesic is monotone decreasing on $(t_1,+\infty)$ and so $\rsln(t_2)<{\rs}_1$. We note that $\rsln$ is also the unique solution on $[t_1,+\infty)$ of the IVP
\begin{eqnarray} \frac{d^2\rs}{dt^2} &=& -\frac{A'(\rs)}{E^2}\nonumber \\
\rs(t_1)&=&{\rs}_1,\qquad \left.\frac{d\rs}{dt}\right|_{t_1}=0.\label{t-geo-ex-ivp-flat}\end{eqnarray}
The initial condition $\rs'(t_1)=0$ fixes $E^2=A({\rs}_1)$. 

We establish existence by considering different values of ${\rs}_2$ as delimited by $\rsln(t_2)$ and by ${\rs}_1\pm(t_2-t_1)$. 

\begin{itemize}
\item[(i)] \textbf{$\rsln(t_2)<{\rs}_2<{\rs}_1+(t_2-t_1)$}. The key here is that the geodesic we seek is initially non-decreasing. Consider the IVP
\begin{eqnarray} \frac{d^2\rs}{dt^2} &=& -\frac{A'(\rs)}{E^2}\nonumber \\
\rs(t_1)&=&{\rs}_1,\qquad \left.\frac{d\rs}{dt}\right|_{t_1}=\sqrt{1-\frac{A({\rs}_1)}{E^2}}.\label{t-geo-ex-ivp2}\end{eqnarray}
For each $E\in[E_*,+\infty)$, a solution of this IVP is uniquely defined on $[t_1,+\infty)$. Furthermore, this solution depends continuously on the parameter $E$, subject to $C^1$ dependence of the coefficients of 
(\ref{t-geo-ex-ivp2}) on $E$, which holds for $E\in[E_*,+\infty)$ (see e.g.\ Section 2.3 of \cite{perko2013differential}). That is, for $t_2>t_1$, the mapping
\[ E:[E_*,+\infty)\to ({\rs}_1-(t_2-t_1),{\rs}_1+(t_2-t_1)): E \mapsto \tilde{\rs}(t_2;E)\]
is continuous, where $\tilde{\rs}(t_2;E)$ is the solution of (\ref{t-geo-ex-ivp2}) evaluated at $t=t_2$. The codomain of this mapping is determined by the null geodesic envelope. In the limit $E\to+\infty$, the solution of (\ref{t-geo-ex-ivp2}) approaches the outgoing null geodesic through $(t_1,{\rs}_1)$, and we have $\lim_{E\to+\infty}\tilde{\rs}(t_2;E)={\rs}_1+(t_2-t_1)$. Similarly, we have $\tilde{\rs}(t_2;E_*)=\rsln(t_2)$. By the continuous dependence result mentioned above, it follows that for all ${\rs}_2\in(\rsln(t_2),{\rs}_1+(t_2-t_1))$, there exists $E_2\in(E_*,+\infty)$ such that $\tilde{\rs}(t_2;E_2) = {\rs}_2$. This gives use the required timelike geodesic connecting $(t_1,{\rs}_1)$ and $(t_2,{\rs}_2)$. 
\item[(ii)] \textbf{${\rs}_1-(t_2-t_1)<{\rs}_2<\rsln(t_2)$}. Existence is proven in this case by applying the argument of case (i) to the IVP
\begin{eqnarray} \frac{d^2\rs}{dt^2} &=& -\frac{A'(\rs)}{E^2}\nonumber \\
\rs(t_1)&=&{\rs}_1,\qquad \left.\frac{d\rs}{dt}\right|_{t_1}=-\sqrt{1-\frac{A({\rs}_1)}{E^2}}.\label{t-geo-ex-ivp3}\end{eqnarray}
\end{itemize}

So we conclude the following: Given any pair of events $(t_1,{\rs}_1)$ and $(t_2,{\rs}_2)$ that satisfy the timelike separation condition $|{\rs}_2-{\rs}_1|<|t_2-t_1|$, there exists a timelike geodesic of $\Mtwo$ that connects these two events. 

We have thus proven in this section  that $\Mtwo$ is a causal domain.
We recall that this has two important implications: first, that the world function $\sigma(x^a,x^{A'})$ is well-defined globally on $\Mtwo$ and second that the retarded Green function $\cal{G}$ is likewise globally valid on $\Mtwo$. 


\section{Geodesic distance and van Vleck determinant in $\Mtwo$} \label{sec:sigma_2} 

In this section, we return to the wave equation on $4-D$ space-time. We point out how the world function and Hadamard form of wave equations on $\Mtwo$ have a role in the $4-D$ analysis, and 
we  calculate numerically  two quantities of relevance.
 These quantities are the world function itself and the van Vleck determinant.
  For the calculation we
  apply some of the results obtained above for geodesics on $\Mtwo$, and we note that Theorem 1 provides the foundation on which these results are built: the calculations below are valid globally on $\Mtwo$. 

By separating the angle variables  in the retarded Green function in  conformal Schwarzschild space-time in the usual way via a multipolar decomposition, we can write
\be \hat{G}_R(x,x') = \frac{1}{4\pi}\sum_{\ell=0}^\infty (2\ell+1)\cgl(x^A,x^{A'})P_\ell(\cos\gamma),\label{con-gret-mode}\ee
where $P_\ell$ are Legendre polynomials and $\cgl$ satisfy the PDE for a Green function on the 2-d conformal space:
\be (P-\lam^2)\cgl=-4\pi\frac{r^2}{f}\delta_2(x^A-x^{A'}),\label{gl-pde}\ee
where
\be P \equiv \square_2 + \frac14\left(1-\frac{8M}{r}\right),\qquad \lam \equiv \ell + \frac 12, \label{pop-def}\ee
and $\square_2$ is the d'Alembertian operator of $\Mtwo$.
In coordinates $(t,\rs)$ it takes the familiar form:
\be  \frac{f}{r^2}\square_2 {\cgl}=  -\pdd{\cgl}{t} +\pdd{\cgl}{\rs} = -4\pi\delta_2(x^A-x^{A'}),\label{gl-pde-coords}\ee
where $\cal{G}$ is a Green function in   $\Mtwo$.

The Hadamard form for the retarded Green function is an analytic expression which is only valid in a normal neighbourhood of a spacetime point.
In $(1+1)$-dimensions, the Hadamard form for the retarded Green function is (e.g.,~\cite{Hadamard,Decanini:Folacci:2008}):
\be{\cal{G}}(x^A,x^{A'})=\theta(t-t')\theta(-\sTCS)U(x^A,x^{A'}). \label{eq:Had form 2d}\ee
The two-point function $U(x^A,x^{A'})$ is a Riemann function for the wave equation \cite{Garabedian}, and may be expanded in the so-called Hadamard series
\be U(x^A,x^{A'})= \sum_{k=0}^{\infty}U_k(x^A,x^{A'}) \sTCS^k. \label{eq:Had-series}\ee
Synge's world function $\sTCS$ satisfies
\be
\nabla_A \sTCS \nabla^A \sTCS=2\sTCS,
\ee
with the initial conditions $\lim_{x'\to x}\sigma(x,x')=0$ and  $\lim_{x'\to x}\nabla_A \nabla_B \sigma(x,x')=g_{AB}(x)$.
The Hadamard coefficients $U_k, k\geq0$, in the $(1+1)$-dimensional space-time  satisfy certain recurrence relations in the form of transport equations along the unique geodesic from $x^A$ to $x^{A'}$~\cite{DeWitt:1960,Friedlander,Decanini:Folacci:2008,Ottewill:2009uj}. 
In particular, in $(1+1)$ dimensions, $U_0$ is the square root of the so-called van Vleck determinant $\Delta=\Delta(x^A,x^{A'})$ and it satisfies:
\be U_0=\Delta^{1/2}\quad \Leftrightarrow \quad \sigma^A\nabla_AU_0=(1-\square_2\sigma)U_0,\quad \lim_{x'\to x}\ U_0=1.
\label{U0-def}\ee
Regularity at $\sigma=0$ fixes constants of integration.
This approach applies for any wave operator of the form $P=\Box_2+v^A\nabla_A + w$, where $v^A$ and $w$ are respectively a vector and a scalar field on $\Mtwo$. Note however, that $U_0$ is defined independently of the wave operator (as of course is $\sigma$). 

In the rest of this section we numerically calculate Synge's world function $\sTCS$ and the van Vleck determinant $\Delta(x,x')$ in the 2-d conformal space $\Mtwo$ 
along the {\it unique} timelike geodesic joining two points in $\Mtwo$. Typically, in applications, one is concerned with causally separated points. For null separations, $\sigma=0$ and so the calculation of $\sigma$ for time-like separations is the issue of most relevance. 


\subsection{Calculation of $\sTCS$ directly from the geodesic equations} \label{subsec:sigma_2}

For timelike geodesics, the world function is given by $\sTCS=-\frac{(\Delta\tau)^2}{2}$ where $\Delta\tau$ is the difference between final and initial values of the proper time
  $\tau$ along the unique timelike geodesic joining two timelike separated points $(t_1,r_1), (t_2,r_2)$. From now on we constrain ourselves to a subcritical ($E<E_c$) timelike  ($\epsilon=-1$) geodesic which starts off at an initial point $(\ta,\ra)$, bounces off the potential barrier at the periapsis $r=\rb$
and then reaches a final point $(\tc,\rc)$, in coordinates $(t,r)$. We note that this is the \textit{least} straightforward case, and incorporates all technicalities that would be encountered in the other cases. 

The calculation of $\sigma((t_1,r_1),(t_2,r_2))$ proceeds in two steps. First, using (\ref{eq:del-tau-def}), we can
evaluate  the proper time taken by a particle with energy $E$ to fall from $r=\ra$ to $r=\rb$. We do this for a range of values of $E$. Then, we (numerically) determine the (unique) value of $E$ that corresponds to this particle departing from $r=\ra$ at coordinate time $t=t_1$ and arriving at $r=\rb$ at time $t=t_2$. The relevant value of $\Delta\tau$ can then be identified from the first step. 

\begin{figure}[h!]
\begin{center}
\includegraphics[width=8cm]{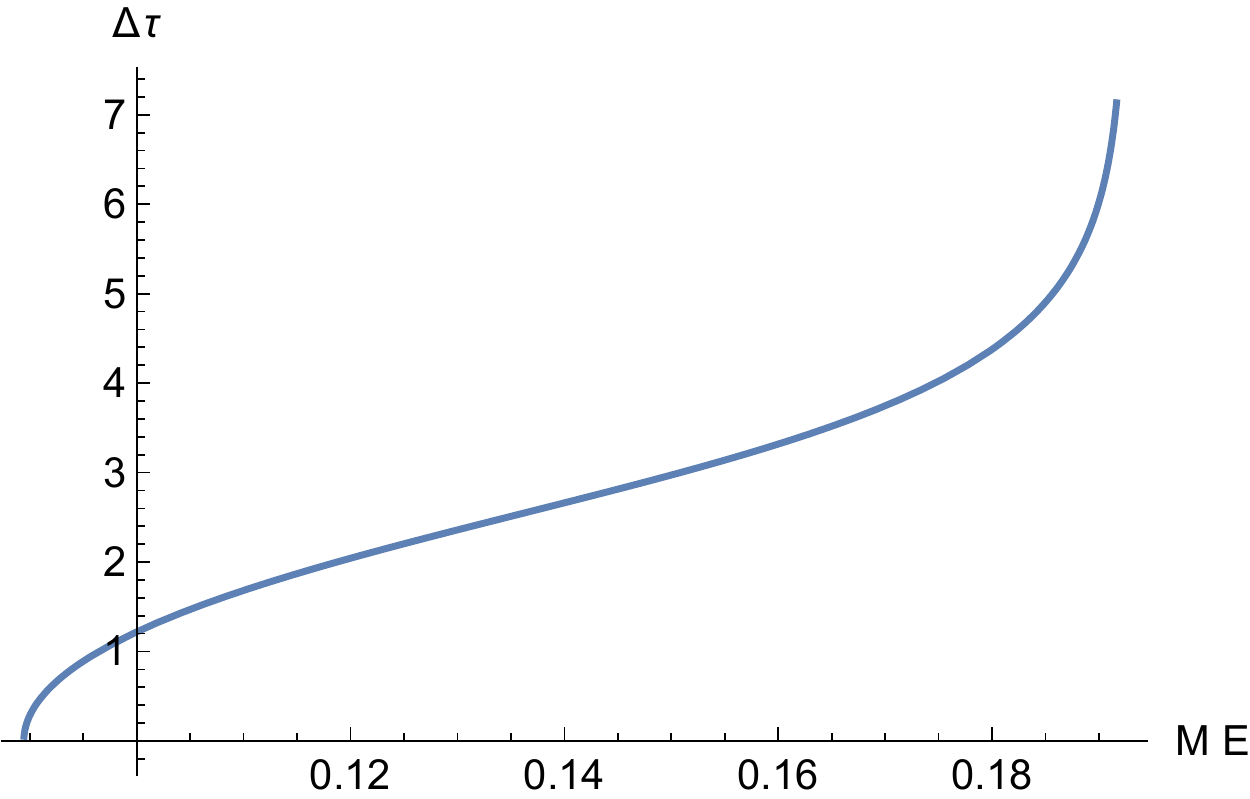}
\includegraphics[width=8cm]{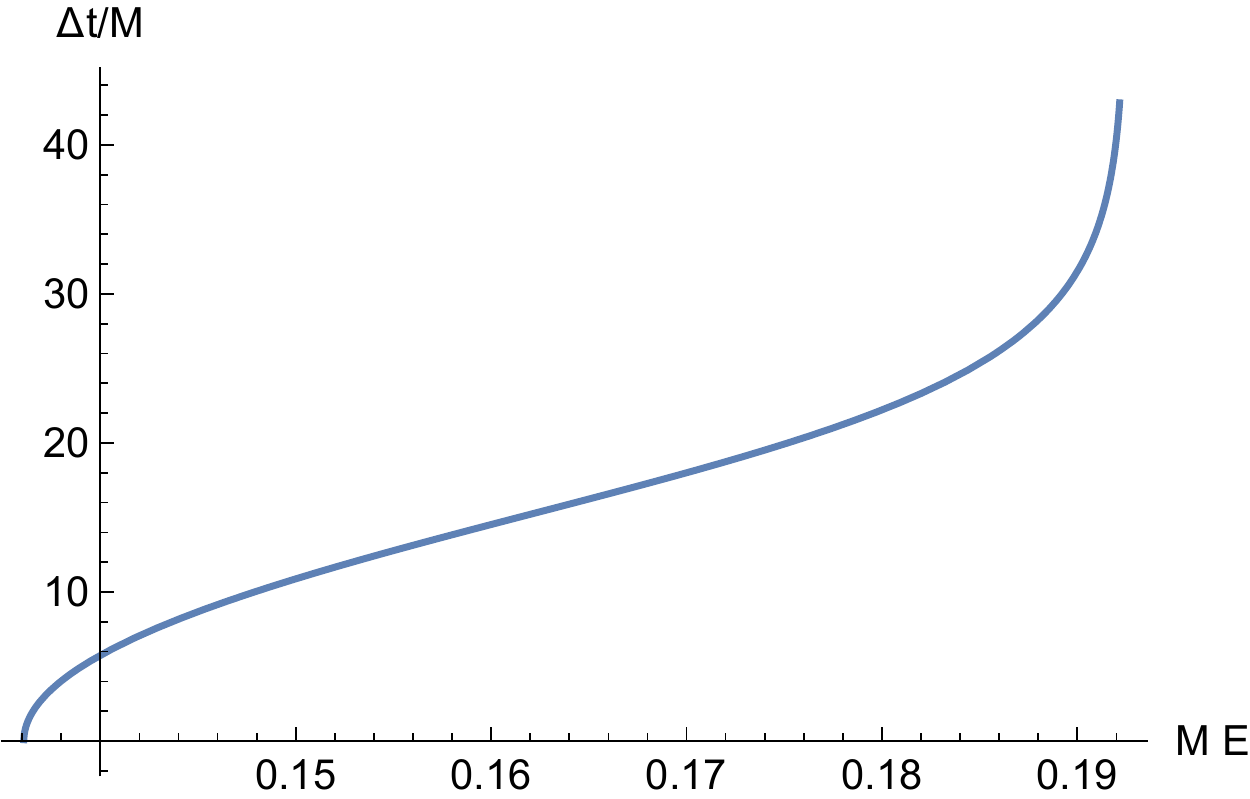}
 \end{center}
\caption{Total interval of 
(a) proper time $\Delta \tau$ from Eq.(\ref{eq:deltau})
and (b) coordinate time $\Delta t$ from Eq.(\ref{t-plus-int2}),
 as functions of energy $E\in (\sqrt{A(\ra)},E_c)$ in the case $\rc=\ra=6M$.
}
\label{fig:Delta tau}
\end{figure}

For convenience, we consider the case that $\rc=\ra$.
Then, in order to calculate the corresponding value of   the total proper time interval  $\Delta\tau$, we calculate Eq.(\ref{eq:deltau}) (where the dependence on $x_0=r_1$ is in the variable $\psi$
-- see Eqs.(\ref{eq:phi-def}) and (\ref{eq:varphi-def}))
and multiply the result by two.
In Fig.\ref{fig:Delta tau}(a) we plot the total proper time interval 
$\Delta\tau$
as a function of the energy for the case $\ra=6M$ (which corresponds to the innermost stable circular orbit
in Schwarzschild space-time).

\begin{figure}[h!]
\begin{center}
\includegraphics[width=8cm]{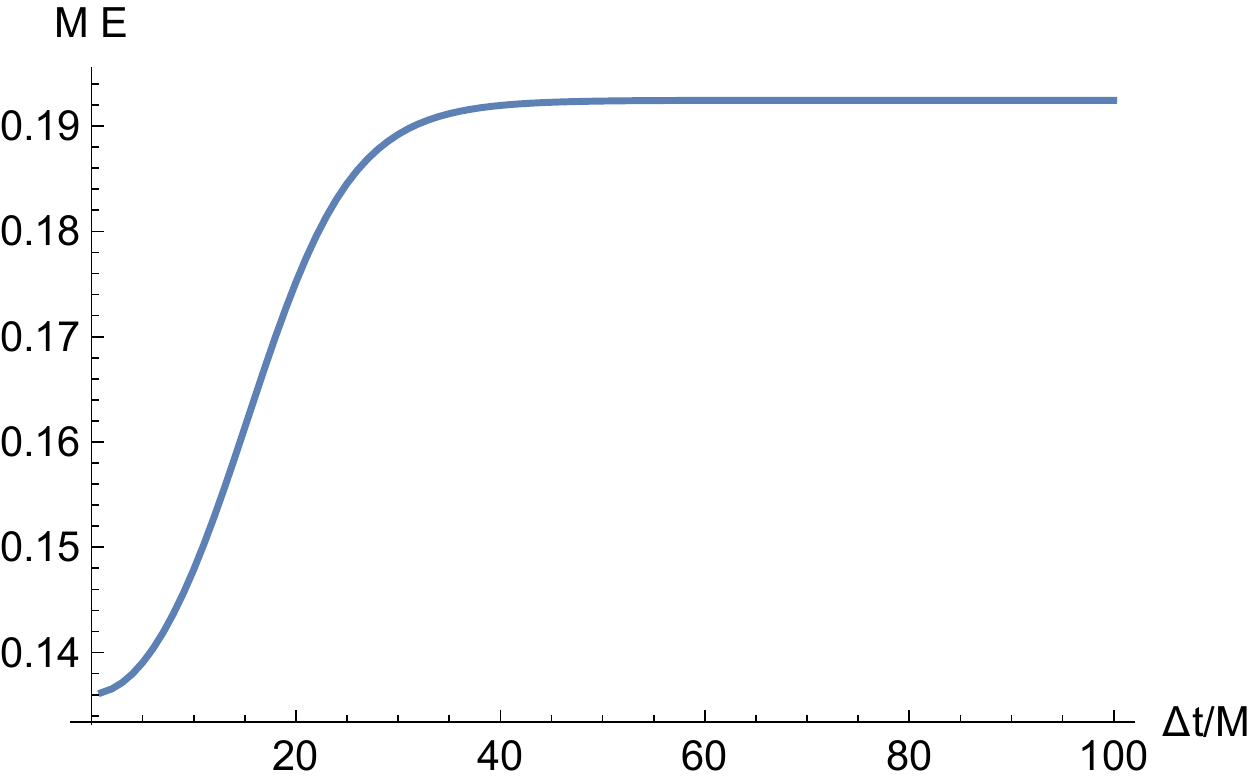}
\includegraphics[width=8cm]{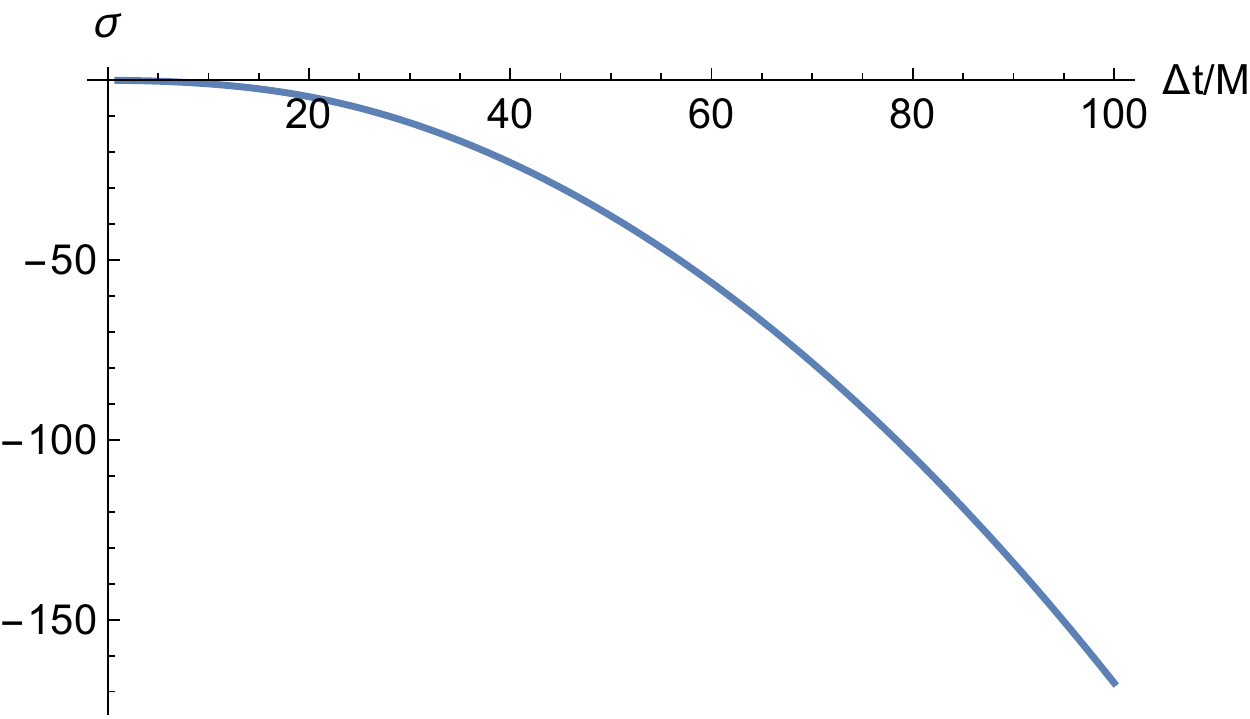}
  \end{center}
\caption{
Various quantities as functions of $\Delta t$ for $\ra=6M$:
(a) $E=E(\Delta t)$ (obtained by zeroizing 
`$\Delta t-2\Delta_1(E)$'
in Eq.(\ref{t-plus-int2})),
and
(b)  $\sTCS=-\tfrac{(\Delta \tau)^2}{2}$ (obtained from  Eq.(\ref{eq:deltau}) using the values of $E$ obtained as in plot (a)).
}
\label{fig:E and tau vs Delta t}
\end{figure}

Similarly for the coordinate time, from Eq.(\ref{t-plus-int2}) we have $\Delta t\equiv \tc-\ta=2\Delta_1(E)$.
This integral can also be solved in terms of various elliptic functions but we do not write in the result as it is not particularly illuminating. 
We plot the result in Fig.\ref{fig:Delta tau}(b).
Eq.(\ref{t-plus-int2}) gives 
$\Delta t=\Delta t(E)$ (for a fixed $\rb$), so in order to find $E=E(\Delta t)$
 we look for a zero of this equation.
We plot the resulting values of $E$ as a function of $\Delta t$ in  Fig.\ref{fig:E and tau vs Delta t}(a).
We can then introduce the values of $E$ in Eq.(\ref{eq:deltau})
 in order to finally obtain the proper time interval  $\Delta \tau$ 
 as a function of  $\Delta t$ for a given $\rc=\ra$. We plot the corrresponding $\sTCS=-\tfrac{(\Delta \tau)^2}{2}$
 as a function of $\Delta t$ in Fig.\ref{fig:E and tau vs Delta t} (b).


\subsection{Calculation of  $\sTCS$  and  $\Delta$ from transport equations} \label{subsec:sigma_2-V0-PDEs}

In this subsection we shall calculate Synge's world function $\sTCS(x,x')$ and the van Vleck determinant $\Delta(x,x')$ in $\Mtwo$ by numerically
solving transport equations that they obey.

The transport equation Eq.(\ref{U0-def}) obeyed by the van Vleck determinant may be simply written as 
\begin{equation} \label{eq:VV}
\frac{d\Delta^{1/2}}{d\tau}=\frac{1}{2 \tau }\left(2-\sigma ^{A}{}_{A}\right)\Delta^{1/2}
\end{equation}
where $\tau$ is an affine parameter along the geodesic joining $x$ and $x'$ and $\sigma ^{A}{}_{B }\equiv \nabla^{A}\nabla_{B}\sigma$.
In order to obtain a transport equation for the derivatives of the world function,
we combine Eqs.(3.11) and (5.2) in~\cite{Ottewill:2009uj}
together with:
\begin{equation}
u^{C}Q^A{}_{B;C}=\frac{dQ^A{}_B}{d\tau}+\Gamma^A_{DC}u^CQ^D{}_B-\Gamma^{C}_{BD}u^{D}Q^A{}_{C},
\end{equation}
 where $u^A$ is the tangent to the geodesic and 
 $Q^{A}{}_{B}\equiv 
\sigma ^{A}{}_{B}
 -\delta^{A }{}_{B}$.
The following equation then follows:
\begin{equation} \label{eq:transp eq Q}
\frac{dQ^{A}{}_B}{d\tau}=u^D Q^A{}_C \Gamma ^C{}_{\text{BD}}-u^D \Gamma ^A{}_{\text{CD}} Q^c{}_B-\frac{1}{\tau }(Q^A{}_C Q^C{}_B+Q^A{}_B)-\tau  R^A{}_{\text{CBD}} u^Cu^D.
\end{equation}

In $\TCSt$, the nonzero Christoffel symbols are
\begin{equation}
\Gamma^{r}_{rr}=\frac{r-M}{r(r-2M)},\quad 
\Gamma^{r}_{tt}=-\frac{6M^2-5Mr+r^2}{r^3},\quad 
\Gamma^{t}_{tr}=\frac{r-3M}{r(r-2M)}=\Gamma^{t}_{rt},
\end{equation}
the nonzero Riemann tensor components are
\begin{equation}
R^{r}{}_{ttr}=-\frac{(r-2M)(r-6M)}{r^4}=-R^{r}{}_{trt},\quad 
R^{t}{}_{rtr}=\frac{r-6M}{r^2(r-2M)}=-R^{t}{}_{rrt},
\end{equation}
and the Ricci scalar is equal to  $\displaystyle R=\frac{12M}{r}-2$.

\begin{figure}[h!]
\begin{center}
\includegraphics[width=8cm]{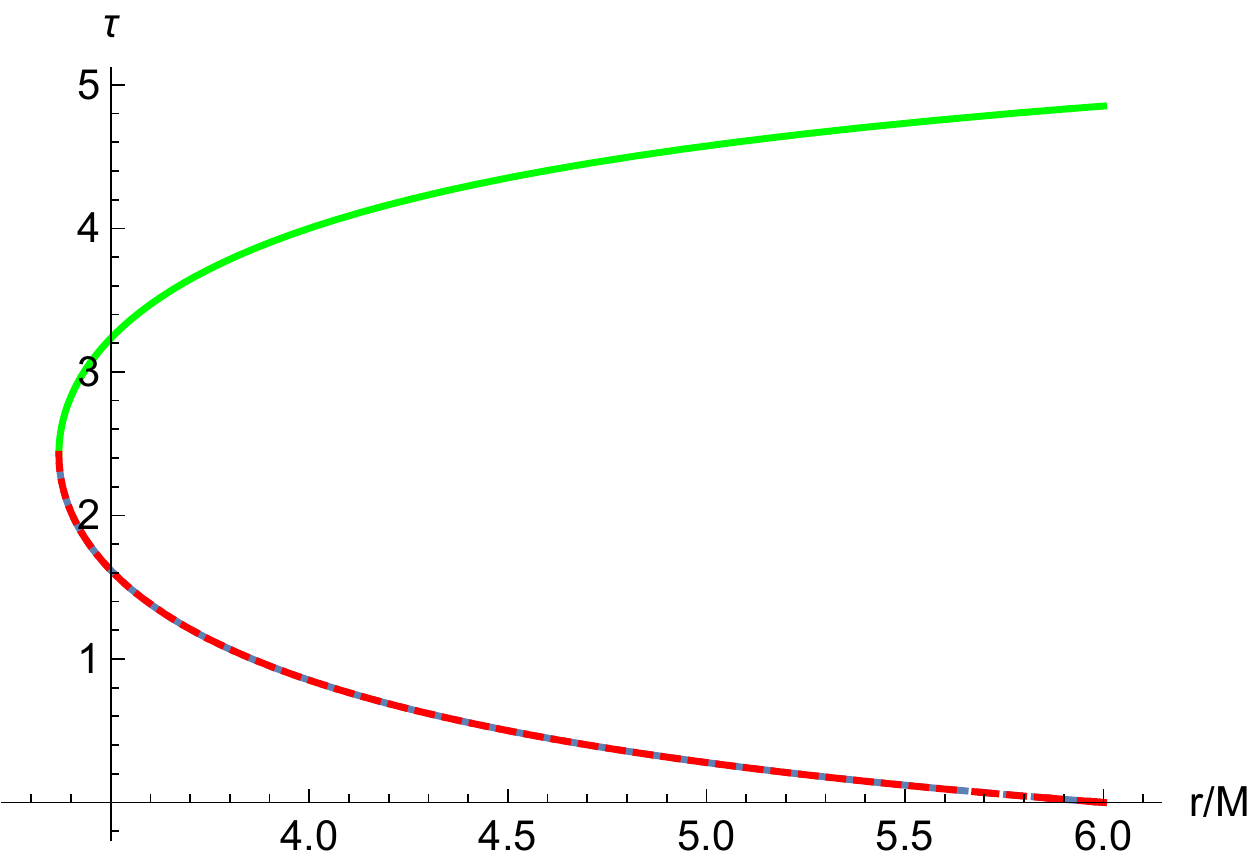}
\includegraphics[width=8cm]{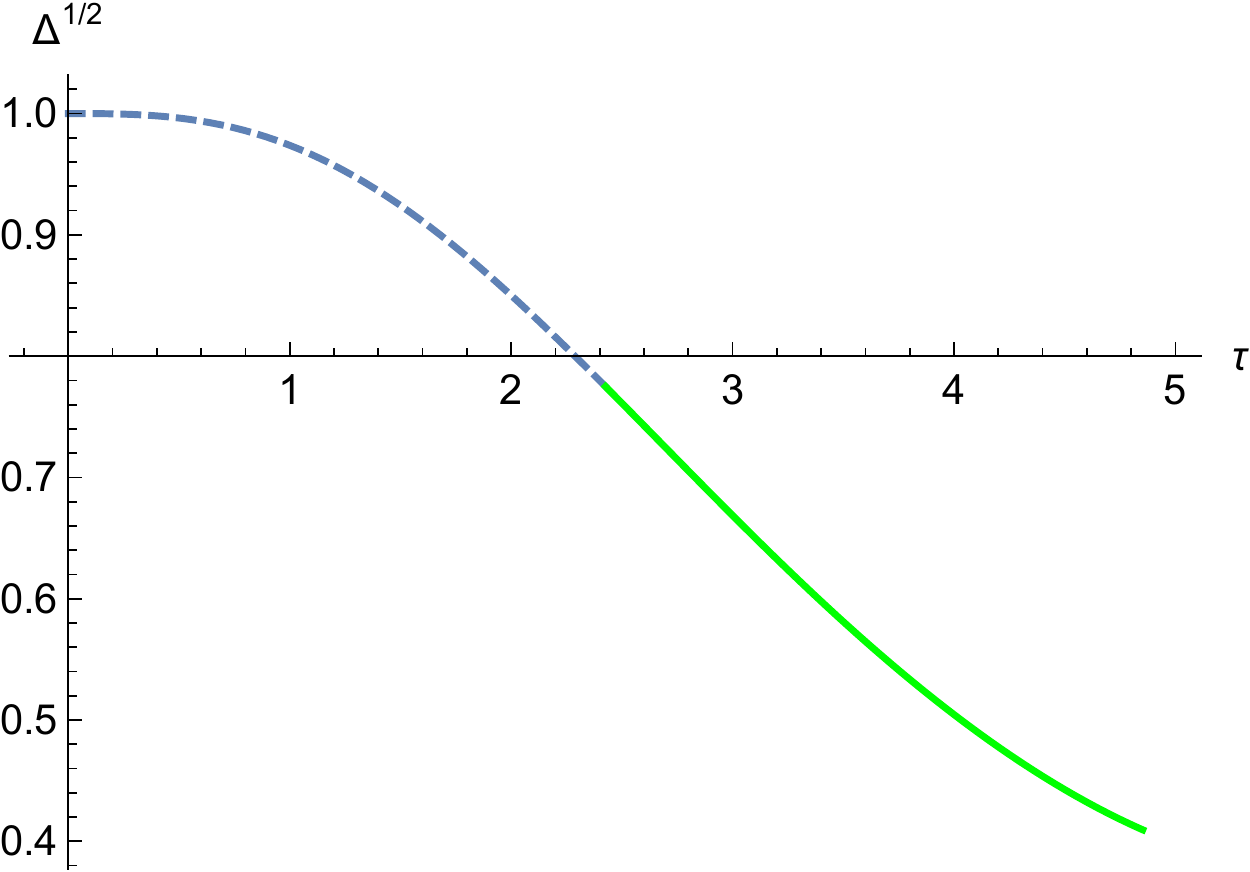}
  \end{center}
\caption{
(a) Proper time, $\tau=\tau(r)$, and (b) square root of the van Vleck determinant, $\Delta^{1/2}=\Delta^{1/2}(\tau)$, along a subcritical timelike geodesic in $\TCSt$.
The geodesic has energy 
$E\approx 0.1892071/M$,
it starts at $r=\ra=6M$, it reaches the periapsis $r=\rb$
and it ends at $r=\ra$.
The results in the dashed blue curve (bottom-half curve in (a) and left-half curve in (b)) and in the straight green curve (top-half curve in (a) and right-half curve in (b))
respectively correspond to the trajectories $r:\ra\to \rb$ and $r:\rb\to \ra$ and
they are obtained using the method described in Sec.\ref{subsec:sigma_2-V0-PDEs}.
The curve in dashed red (overlapping with the dashed blue curve) in (a) is obtained using Eq.(\ref{eq:deltau}) (with the initial value $x_0=r_1$ being replaced by a value of $r$ along the geodesic).
}
\label{fig: tau and VV}
\end{figure}

We can solve Eqs.(\ref{eq:VV}) and (\ref{eq:transp eq Q}) simultaneously by using the code in~\cite{Hadamard-WKB-Code} in
order to obtain $\tau=\tau(r)$ and $\Delta^{1/2}=\Delta^{1/2}(\tau)$ along a specific timelike geodesic in $\TCSt$.
We plot these results in Fig.\ref{fig: tau and VV}.
We checked that the plot of $\tau=\tau(r)$ agrees with the calculation of the proper time using Eq.(\ref{eq:deltau}) (with the initial value $x_0=r_1$ being replaced by a value of $r$ along the geodesic), 
thus serving as a validation of both methods.

 In Fig.\ref{fig:sigma_2,tau,VV Schw} we plot the final values (i.e., upon return to $r=\ra$) of $\sTCS=-(\Delta \tau)^2/2$ and of  $U_0=\Delta^{1/2}$  calculated as above for a geodesic in $\TCSt$ 
 for specific values of $\Delta t$ and of $\ra=\rc$.
 These final values are the relevant ones for the expression for the Green function in Schwarzschild spacetime that we use in~\cite{casals-nolan-global-hadamard}.
  
\begin{figure}[h!]
\begin{center}
\includegraphics[width=8cm]{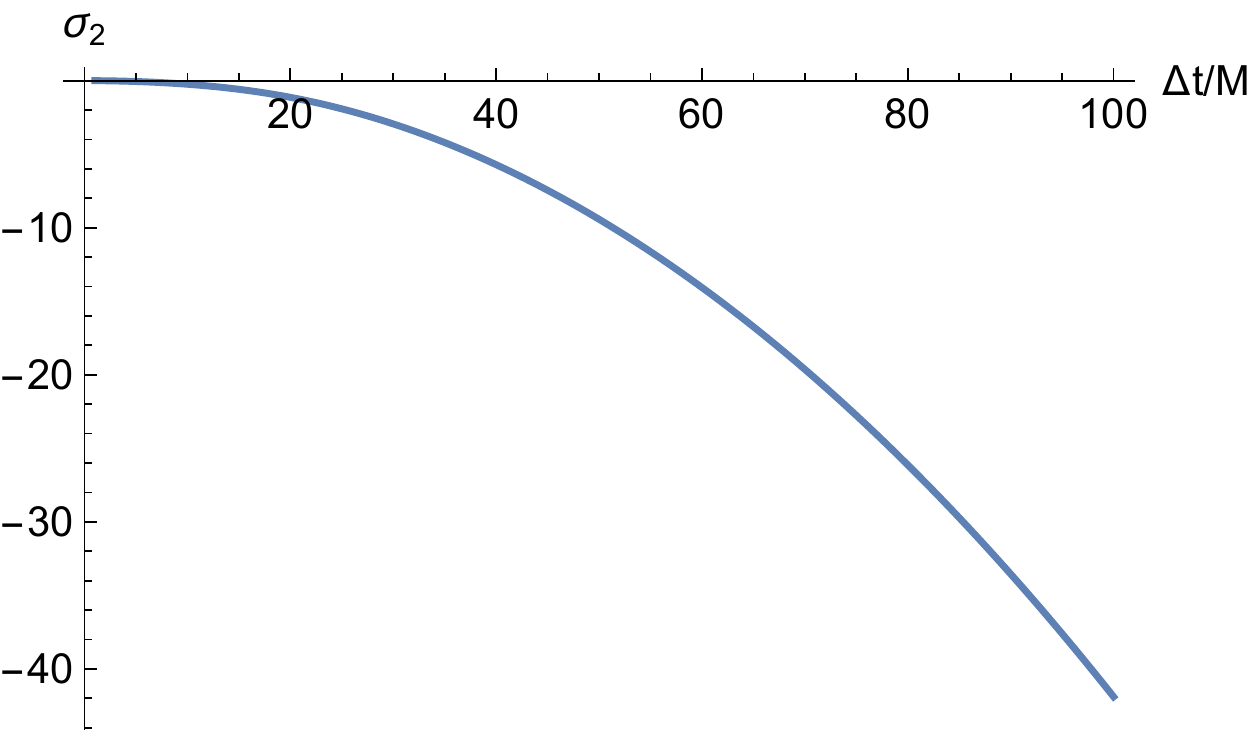}
\includegraphics[width=8cm]{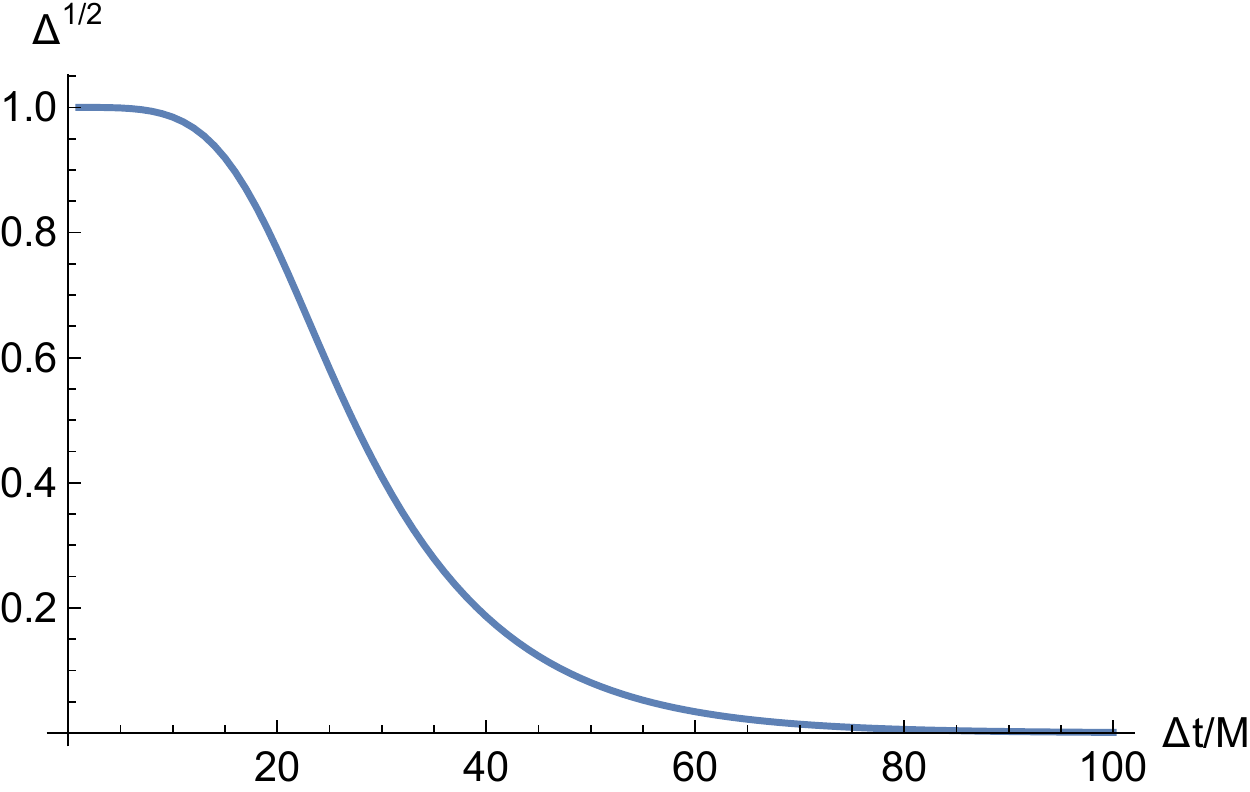}
  \end{center}
\caption{
Final values  (i.e., at the return at $r=\ra$) of: 
(a) the world function $\sTCS=-(\Delta \tau)^2/2$  and (b) the square root   of the van Vleck determinant, $\Delta^{1/2}$, along the only (subcritical and timelike) geodesic in $\TCSt$ 
 as functions of $\Delta t$ for $\ra=\rc=6M$.
}
\label{fig:sigma_2,tau,VV Schw}
\end{figure}


\section{Conclusions} \label{sec:conclusions}

The main motivation for this paper is ultimately the study of the self-force problem in General Relativity (although our results may also be useful for other problems, such as in the calculation of the response of a quantum ``particle detector"). Approaches to this problem that rely on the study of Green functions in black hole spacetimes have yielded significant geometrical insights. Our particular focus (see e.g.\ \cite{Casals:2012px}) is to determine global information on Green functions in black hole spacetimes. In the case of the exterior Schwarzschild spacetime, as seen in Section~\ref{sec:sigma_2}, this is aided by a thorough understanding of Green functions on the spacetime $\Mtwo$. Thus $\Mtwo$  is a space-time of physical relevance: it appears naturally in the wave equation in Schwarzschild space-time when applying a technically-convenient rescaling of the scalar field. Exploiting the natural occurrence of $\Mtwo$ will be greatly aided by the geometrical property we have established here - i.e.\ that $\Mtwo$ is a causal domain. 

On the one hand, such a proof is of theoretical interest since being a causal domain is a rather rare property of space-times. For example, black hole space-times, where closed null geodesics are a ubiquitous feature \cite{stuchlik2000equatorial}, are not causal domains. Likewise, plane-wave spacetimes generically contain pairs of points connected by multiple geodesics, and so are not causal domains \cite{harte2012caustics}. De Sitter space-time contains pairs of points not connected by any geodesic \cite{Hawking:Ellis}. Thus geodesic convexity fails, and de Sitter space-time is not a causal domain. Anti-de Sitter shares this feature, and furthermore, there exist pairs of points $p,q$ for which $J^+(p)\cap J^-(q)$ is non-empty and non-compact \cite{bar2009quantum}: thus anti-de Sitter space-time fails both criteria required for a causal domain. (However, it should be noted that this failure to be a causal domain does not prevent one from constructing global Green functions on de Sitter space-time \cite{Birrell:Davies}.).

From a mathematical point of view, the property of being a causal domain is unstable - at least in the case of $\Mtwo$. To see this, consider perturbing the metric so that the effective potential shown in Figure \ref{fig:Potential} has a local minimum. This local minimum would act as a potential well, trapping particles in a particular range of energies. Two particles with different energies, both confined to this potential well, will meet repeatedly, providing examples of pairs of points connected by multiple geodesics. The perturbation of the potential, and so of the metric tensor (and curvature tensor) may be arbitrarily small, and so a small perturbation of the space-time violates its causal domain nature. 

On the other hand, the property of being a causal domain is also of practical interest. The reason is that, being a causal domain implies that the world function is defined \textit{globally} and that the analytic Hadamard form  for the retarded Green function is valid for any arbitrary pair of  points in $\Mtwo$. This feature is exploited in a separate study~\cite{casals-nolan-global-hadamard}, where we derive global properties of the Green function in Schwarzschild space-time by expressing it in terms of the world function and the Hadamard coefficients (such as the van Vleck determinant) in $\Mtwo$. 

Finally, we note that the focus on $\Mtwo$ allows us to understand more clearly the global structure of Schwarzschild space-time in the following way. Consider null geodesics on Schwarzschild space-time. By conformal invariance of null geodesics, these are also null geodesics on the conformal Schwarzschild spacetime. The motion then projects onto geodesic motion on $\mathbb{S}_2$ and geodesic motion on $\Mtwo$. Suppose that there are multiple null geodesics connecting the $4-D$ points $x^\alpha=(t_1,r_1,\theta_1,\phi_1)$ and $x^{\alpha'}=(t_2,r_2,\theta_2,\phi_2)$. Then the causal domain property of $\Mtwo$ (in particular its geodesic convexity) implies that the projection of these two geodesics onto $\Mtwo$ yields the same geodesic in $\Mtwo$ (but parametrized by different affine parameters). The projections into $\mathbb{S}_2$ yield geodesics on $\mathbb{S}_2$ that can differ only in terms of the particular great-circular path that they follow. Thus null geodesics that connect the same points of Schwarzschild space-time trace the same path in $t-r$ (or $t-\rs$) space. As exploited in~\cite{casals-nolan-global-hadamard}, this enables us to define a globally valid 2-point function on conformal Schwarzschild space-time that provides a global version of the world function.


\begin{acknowledgments}
We are thankful to Barry Wardell for useful discussions.
\end{acknowledgments}

\end{document}